\documentclass[aps,prb,preprint,groupedaddress]{revtex4-2}

\setcitestyle{numbers,square}

\usepackage{hyperref}
\usepackage{graphicx,rotating}
\usepackage{dcolumn}
\usepackage{bm}
\usepackage{epsfig}
\usepackage{longtable}
\usepackage{color}
\usepackage[T1]{fontenc}
\usepackage{mathtools}
\usepackage{amsmath}

\begin{document}

\title{Nature of light absorption in Eu$^{3+}$ doped tetragonal ZrO$_2$ with oxygen vacancies,
bulk and [001] surface: charge-transfer excitons obtained from {\it ab initio} Bethe-Salpeter equation}

\author{Małgorzata Wierzbowska}
\email[]{malwi45@gmail.com}
\affiliation{Institute of High Pressure Physics Polish Academy of Sciences, Soko\l{}owska 29/37, 01-142 Warsaw,
Poland}

\date{\today}

\begin{abstract}
Light absorption spectra were calculated using the {\it ab initio} Bethe--Salpeter equation ({\it ai}-BSE) for tetragonal ZrO$_2$ 
doped with single and double Eu$^{3+}$ ions in the bulk and on the [001] surface, with single and double oxygen vacancies. 
Both the Hubbard $U$ correction and spin--orbit coupling were included. Several impurity coordination symmetries were analyzed 
to investigate the excitonic character of the main absorption peaks. 
In none of the studied cases the $f-f$ excitations were observed.
Instead, the holes of the charge-transfer excitons were delocalized over the oxygen atoms in all cases.
The excited electrons were localized at Eu for the low-energy peaks and Zr for dipole transitions across the fundamental band gap. 
These findings challenge the common assumption that optical transitions in all rare-earth doped systems occur solely between 
impurity energy levels. Therefore, spectra cannot be reliably described using only free-ion models, such as the Judd--Ofelt theory, 
even when crystal-field splitting, spin--orbit coupling, and thermal corrections are included. 
Excitation of the deeply bound valence states, such as those of Eu-$4f$ character, 
would require a far-UV and 
extremely intense light source to cause a massive population-inversion of strongly hybridized bands.
\end{abstract}

\maketitle

\section{Introduction}

Rare-earth doped materials find broad applications in optoelectronics. 
White LEDs, for example, are doped with both Eu$^{3+}$ and Eu$^{2+}$ to achieve intense red and blue emission \cite{Eu-LED,blue-red,Eu2BaO}. 
UV-excited phosphors provide visible light emission for cathode-ray tubes, mercury lamps, and electroluminescent displays \cite{UV-devices}. 
Increasing the concentration of Eu$^{3+}$ can enhance luminescence quantum efficiency, even if UV absorption decreases with doping \cite{UV-sensor}. 
In medicine, rare-earth doped materials are used for fluorescence imaging, drug delivery, and image-guided therapy \cite{medicine}, 
while Eu is employed in shielding glasses for nuclear radiation protection \cite{app-radiation}. 
Eu$^{3+}$ serves as a spectroscopic probe to analyze luminescence spectra and characterize local site symmetries in the oxide glass or crystal 
\cite{envir-probe,cryst-struc}, whereas Er$^{3+}$, Tm$^{3+}$, and Yb$^{3+}$ are used for light upconversion \cite{upconversion}. 
Lanthanides in molecular systems enable luminescent probes and sensors \cite{f-mol}, and in organic compounds they exhibit 
circularly polarized luminescence applied in security inks \cite{appl-inks}. 
Dual-color emission of Eu$^{3+}$-doped ZrO$_2$ nanoparticles under 254 nm (4.88 eV) and 365 nm (3.40 eV) excitation is used for 
anti-counterfeiting \cite{anti-counterfeiting}. 
Rare-earth doped single-crystal (Er:CaF$_2$) \cite{cryst-fiber} and glass fiber lasers \cite{fibre-nasi} allow both short-pulse 
and continuous-wave operation \cite{high-power}. 
Device models often incorporate impurity-level diagrams, exciton relaxation times, and absorption/emission cross sections \cite{laser-model}, 
highlighting the need to understand exciton dynamics and localization.

The most common phenomenological approach to describe light absorption and emission in rare-earth doped systems is the free-ion model, 
known as the Judd--Ofelt theory (JOT), which considers only $f-f$ optical transitions 
\cite{JO-1,JO-2,JO-ff,Eu-big,model-1,Brik-model,from-Witek,morphology,lumin-ff-ZrO2,O-blue}. 
Rare-earth $f$ electrons are strongly correlated, subject to significant spin--orbit coupling (SOC), and sensitive to their local environment. 
The JOT includes crystal-field, SOC, and thermal effects in a parametric manner. 
Traditionally, it is assumed that $f-f$ dipole transitions dominate the absorption and emission spectra of lanthanide ions embedded 
in oxide crystals \cite{tZrO2-orange,oxide-2,lumin-ff-ZrO2,from-Witek,morphology}, nanoparticles \cite{nano-1,nano-2}, 
and glass matrices \cite{glass-1,glass-2,glass-3,glass-2024}. 
Deviations of observed peaks from theoretical $f-f$ positions have been attributed either to the combination of multiple 
$f$-shell transitions \cite{glass-1} or shifts induced by geometrical distortions affecting the crystal-field symmetry, 
with contributions from electron and hole trap states \cite{glass-2}. 
A comprehensive discussion of Eu$^{3+}$ $f-f$ transitions within the JOT framework is given in Ref. \cite{Eu-big}.
A key limitation of the JOT is the neglect of excitations between impurity and host states. 
Some $f$-shell levels may be deeply buried below the Fermi level requiring a deep UV light for their excitation. 
In addition, a strong hybridization of the impurity states with the neighboring atoms calls for a larger number of photons in the volume.
Occupation of excited states can, in turn, modify their energetic positions. Overlaps between measured peaks and theoretical 
$f-f$ transitions may be coincidental, arising from the localized states due to the defects such as oxygen vacancies, impurities, 
or hydroxyl groups \cite{defects}. Moreover, the optical transitions localized at the impurity are relatively temperature-insensitive, 
whereas glass ceramics are highly temperature-sensitive \cite{temp-1,temp-3} and are used in optical temperature sensors \cite{opt-temp}. 
This observation indicates that many optical transitions have an inter-atomic rather than purely intra-site character, 
as inter-atomic transitions are strongly coupled to phonon and thus to temperature.

Simpler phenomenological models rely on knowledge of: (i) the fundamental band gap via Tauc's approach \cite{Tauc}, 
(ii) the optical band gap (absorption edge) through theabsorption spectrum fitting (ASF) \cite{ASF-1,ASF-2}, 
(iii) the exciton binding energy from the Elliott fit \cite{Elliott} or hydrogenic exciton models (HEM) \cite{HEM-1,HEM-2}, 
(iv) the incident photon energies near the absorption edge, 
and (v) the electron--phonon and exciton--phonon coupling by the Urbach law \cite{Tauc}. 
A brief overview of these models is provided in Ref. \cite{glass-1}.

On the other hand, the very advanced {\it ab initio} approaches allow a description of the optical dipole transitions and mapping of excitons onto
the atomic lattice \cite{yambo-1,yambo-2}; it is feasible for crystals, nanostructures and molecules. 
The glass systems cannot be treated as periodic and the embedded cluster methods must be employed \cite{cluster}.

\begin{figure*}
\includegraphics[scale=0.23,angle=0.0]{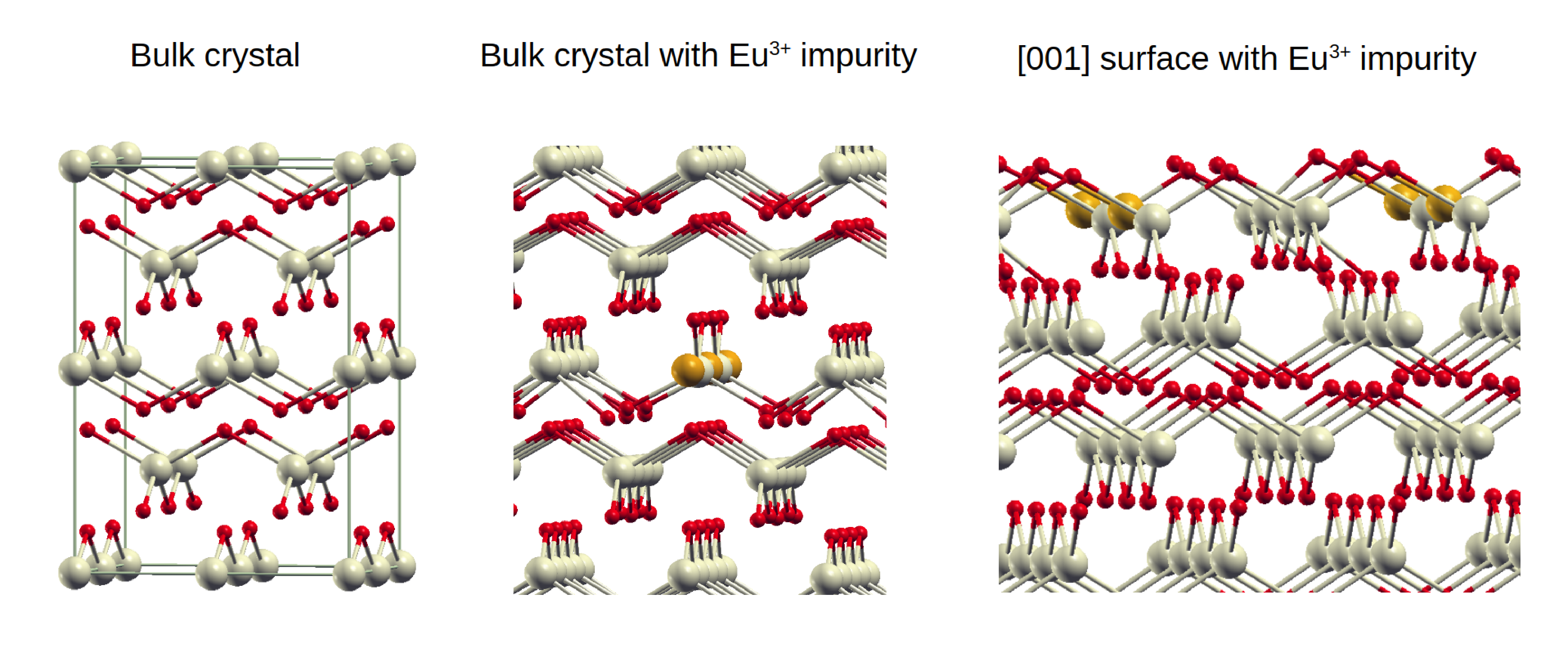}
\caption{\label{fig1} Symmetry of the atomic environment in the tetragonal ZrO$_2$ bulk crystal and [001] surface.
Color legend: Zr - cream, O - red, Eu - orange. }
\end{figure*}
In this work, the tetragonal ZrO$_2$ crystal (t-ZrO$_2$) 
with oxygen vacancies is chosen as a model host for Eu$^{3+}$ ions. 
This choice reflects the environmental similarities between impurities in t-ZrO$_2$ and oxide glasses \cite{MD-glass,ZrO2-geom}, 
which may result in the comparable spectral characteristics for a few reasons:
(i) the average Eu--O bond length is $\sim$2.4--2.5 \AA\ in oxide glasses and 2.09--2.43 \AA\ in t-ZrO$_2$,
(ii) the average impurity coordination number is 4.5--4.9 in oxide glasses 
compared with 8 in the bulk and 6 at the surface of t-ZrO$_2$, while an introduction of oxygen vacancies reduces these numbers,
(iii) in the both systems, impurities do not interact for a longer distance (reports 
about the existence of luminescent probes of local structure were published for glasses \cite{envir-probe} and crystals \cite{cryst-struc}).  
Although many local similarities between glasses and crystals, especially with respect to the first atomic shell around impurity,
the differences are mainly in the statistical character of distortions and disorder in amorphous systems, and its lack in crystals.
If we wanted to mimic the spectral properties of the oxide glass by taking a weighted sum of the spectra for many defects in a crystal, 
then we should also average over the spectra of given impurity in many oxide crystals or crystal alloys
containing cations of the chosen oxide glass. 
Nevertheless, the first step analysis is done in this work.

Figure~1 illustrates the atomic environment in t-ZrO$_2$: each O atom forms four bonds with Zr, two of which are longer, 
while each Zr atom forms eight bonds with O, four of which are longer. At the [001] surface, two of the longer Zr--O bonds 
(to the second-neighbor oxygen layer) are broken, resulting in four shorter and two longer bonds per Zr atom. 
Oxygen vacancies, both neutral and positively charged, have been investigated theoretically and experimentally, 
accounting for the plausible existence of Zr$^{3+}$ \cite{O-charge}, and the negative charge transfer from the vacancy 
to Zr$^{4+}$ was not observed \cite{O-charge}. 
Vacancies introduce the trap states contributing to blue luminescence \cite{O-trap,O-blue} 
and enhance visible-light photocatalytic performance \cite{O-photocatalysis}.

ZrO$_2$ exists in three polymorphous crystals: monoclinic (stable at low temperature), tetragonal (stable above $\sim$1173$^{\circ}$C), 
and cubic (stable above $\sim$2370$^{\circ}$C) \cite{geom-1,geom-2,geom-3}. 
The room-temperature tetragonal phase (t-ZrO$_2$), which is a meta-stable in the pure crystal, can be stabilized by doping with 
Y$_2$O$_3$ \cite{Y2O3-tZrO2} or with Sc and Ce oxides \cite{Sc-Ce-Y}. 
The stabilization mechanism with yttrium was theoretically studied including the surface effects \cite{Fabris,Ricca}, 
and was experimentally observed in other Eu-doped oxides such as heavy scintillators Lu$_2$O$_3$:Eu and 
HfO$_2$:Eu \cite{heavy-scintillator}. Increased concentration of Eu further stabilized t-ZrO$_2$ \cite{nano-1}.
Experimental band gaps for c-ZrO$_2$, t-ZrO$_2$, and m-ZrO$_2$ are 6.10 eV, 5.83 eV, and 5.78 eV, respectively \cite{ZrO2-geom}. 
Defect states within the fundamental band gap of t-ZrO$_2$:Eu$^{3+}$ led to stable orange-red luminescence \cite{tZrO2-orange}, 
while oxygen vacancies reduced the band gap to $\sim$1.5 eV useful for solar photovoltaic applications \cite{solar-PV}. 
Nanoparticles of t-ZrO$_2$ were synthesized via sol--gel \cite{sol-gel} and forced hydrolysis of zirconyl chloride \cite{hydrolysis}. 
Recently, the optical properties of yttrium-doped and Fe-codoped t-ZrO$_2$ crystals were reported using the similar  approaches 
to this work; however, those studies did not include the rare-earth dopants 
and investigated properties differed \cite{Fe-ZrO2}.

\begin{figure*}
\includegraphics[scale=0.22,angle=0.0]{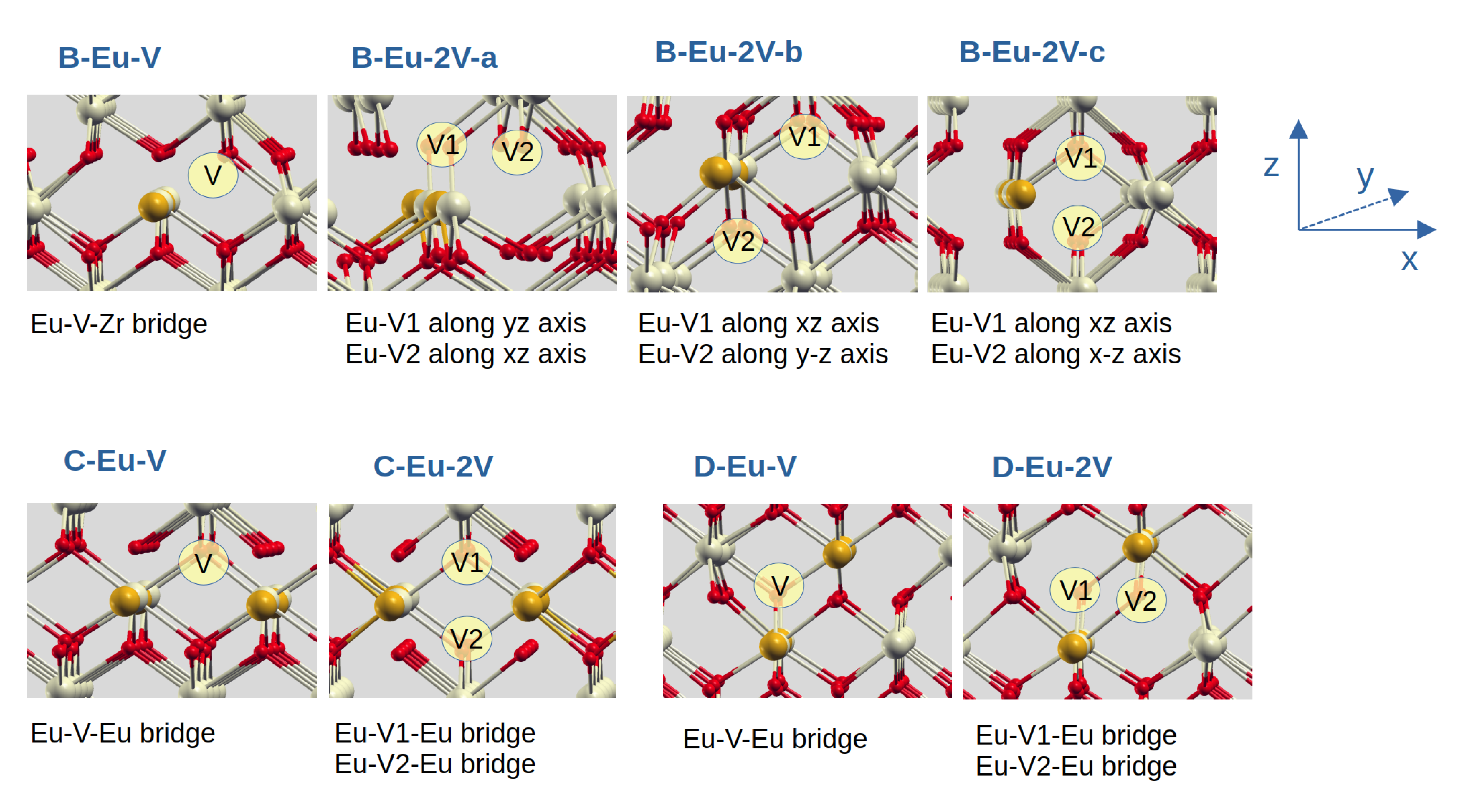}
\caption{\label{fig2} Position of oxygen vacancies in the t-ZrO$_2$ crystal with single Eu$^{3+}$ impurity
	('B-Eu-V', 'B-Eu-2V-a,b,c'), and two Eu$^{3+}$ impurities in the same Zr-layer ('C-Eu-V', 'C-Eu-2V') and 
	neighboring Zr-layers ('D-Eu-V', 'D-Eu-2V').}
\end{figure*}
In this work, the light absorption spectra have been calculated using {\it ab initio} Bethe--Salpeter equation ({\it ai}-BSE) 
for several configurations of Eu$^{3+}$-doped t-ZrO$_2$: 
a single Eu$^{3+}$ ion substituting Zr$^{4+}$ in the bulk and no oxygen vacancy (‘B-Eu’); 
the same configuration with one oxygen vacancy (V$_\mathrm{O}$) near Eu$^{3+}$ (‘B-Eu-V’);
the same configuration with two V$_\mathrm{O}$ near Eu$^{3+}$ placed in the atomic positions shifted along the $xz$ and $yz$ axes (‘B-Eu-2V-a’), 
along the $xz$ and $y\bar{z}$ axes (‘B-Eu-2V-b’), and along the $xz$ and $x\bar{z}$ axes (‘B-Eu-2V-c’);
two Eu$^{3+}$ ions substituting Zr$^{4+}$ in the same layer and no oxygen vacancy (‘C-Eu’);
the same configuration with one V$_\mathrm{O}$ between two Eu$^{3+}$ ions (‘C-Eu-V’) and two V$_\mathrm{O}$ (‘C-Eu-2V’);
two Eu$^{3+}$ ions placed at Zr$^{4+}$ sites in the neighboring layers (‘D-Eu’), with one V$_\mathrm{O}$ (‘D-Eu-V’), and two V$_\mathrm{O}$ (‘D-Eu-2V’); 
[001] surface with a single Eu$^{3+}$ ion substituting Zr$^{4+}$ and no oxygen vacancy (‘S-Eu’);
the same with one V$_\mathrm{O}$ near Eu$^{3+}$ and the neighboring oxygen atom moved to a top of the surface Zr$^{4+}$ (‘S-Eu-V’). 
The atomic structures of bulk cases are shown in Fig.~2, while the S-Eu-V configuration is presented in Fig.~3. 
Cases with two Eu$^{3+}$ ions have been studied for two types of local spin ordering: ferromagnetic (FM) and antiferromagnetic (AF). 
Each V$_\mathrm{O}$ introduces the charge +2 (after removal of O$^{2-}$) and each Eu$^{3+}$ ion adds the charge +1 (replacing Zr$^{4+}$). 
Consequently, a neutral unit cell of t-ZrO$_2$:Eu$^{3+}$ contains one oxygen vacancy (+2) per two Eu$^{3+}$ ions.
\begin{figure}
\includegraphics[scale=0.16,angle=0.0]{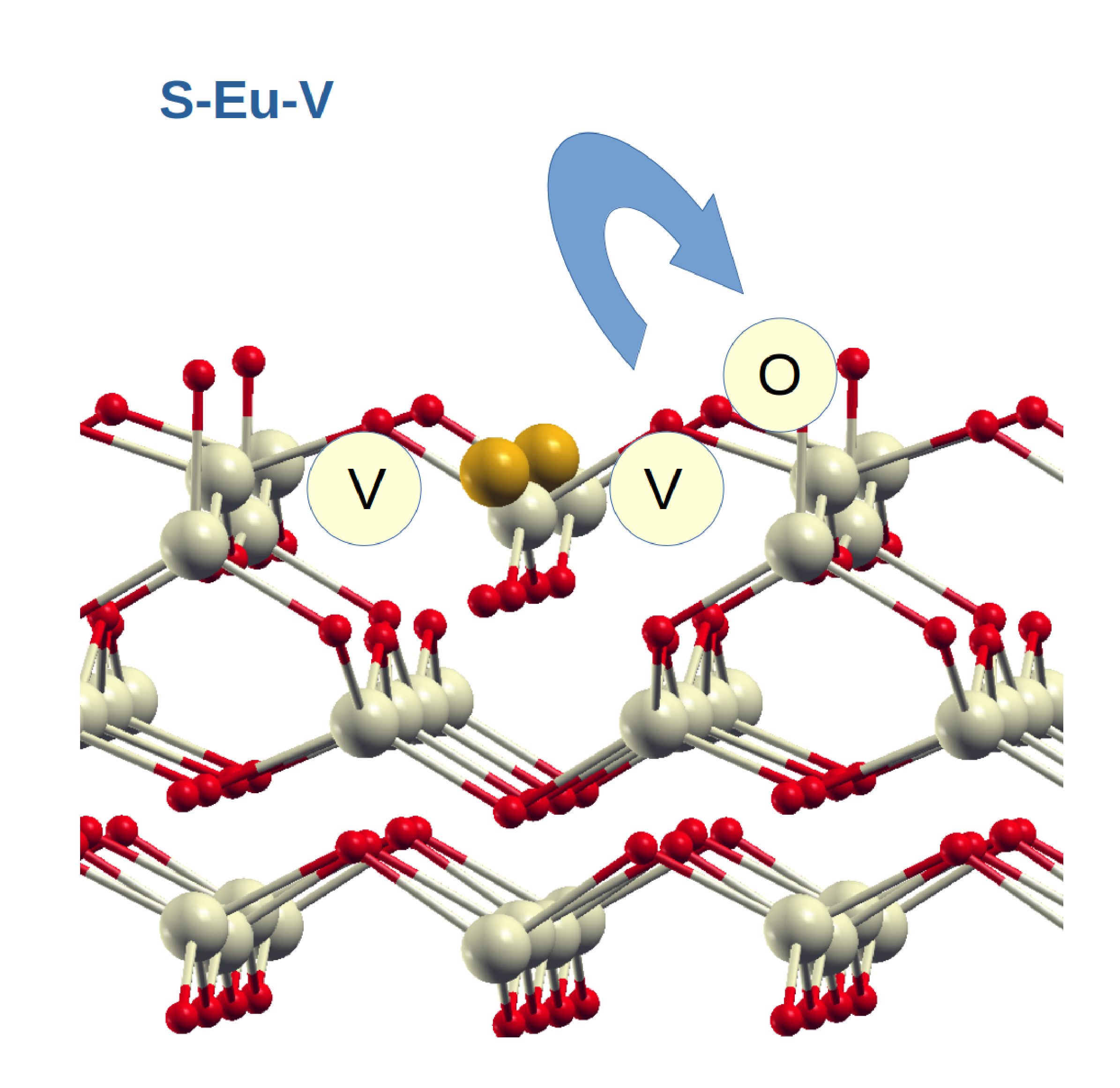}
	\caption{\label{fig3} Oxygen vacancy in top layer of the t-ZrO$_2$ [001] surface with single Eu$^{3+}$ impurity ('S-Eu-V').
	Blue arrow indicates a jump of the oxygen adjacent to Eu$^{3+}$ on the Zr of top layer.}
\end{figure}
Additionally, the set of studied systems is extended by a case which corresponds to 'B-Eu' with the Eu$^{3+}$ ion replaced by Eu$^{2+}$ 
('B-Eu-e'). The Eu$^{2+}$ ion, which can be introduced into ZrO$_2$ from EuF$_2$, was previously investigated 
in Y$_3$Al$_5$O$_{12}$ (YAG) phosphors \cite{Eu2YAG}, and used in various applications of inorganic scintillators: 
high-energy physics, positron emission tomography and $\gamma$-ray detection, where the high light-yield, good energy resolution 
and fast response time were required \cite{Eu2BaBrI}. 
Eu$^{2+}$ also stabilized the organic silicate glasses \cite{organic-1} which were used for interconnectors in CMOS technology.

For all cases studied in this work, the band structure composition and band-to-band transition energies are analyzed in the context of 
the calculated absorption spectra. The strongest dipole transitions associated with the main peaks and 
spatial localization of exciton holes and electrons are discussed for the both real and reciprocal space.

\section{Methods}

The structural relaxation and electronic band structures were calculated with the density functional theory (DFT) \cite{Kohn,Sham} 
using Quantum ESPRESSO package \cite{QE}. The local density approximation (LDA) for the exchange--correlation functional was employed \cite{PZ}. 
Norm-conserving pseudopotentials were used with a plane-wave energy cutoff of 80 Ry, which yields converged energy levels. 
The first Brillouin zone (BZ) was sampled using a uniform Monkhorst--Pack grid \cite{MP} with an 8$\times$8$\times$8 $k$-point mesh 
for bulk systems containing 48 atoms in the elementary cell and an 8$\times$8$\times$1 mesh for surface slabs consisting of 72 atoms 
separated by a vacuum region of 45 \AA. Atomic positions were relaxed using the Broyden--Fletcher--Goldfarb--Shanno (BFGS) algorithm \cite{BFGS}. 
Charged impurities (additional electrons or holes in the elementary cell) were treated using a compensating uniform background charge \cite{charge}. 
Interpolation of the band structures was performed using maximally localized Wannier functions (MLWFs) \cite{wan1,wan2} 
implemented in the wannier90 code \cite{wan90}. 
The high-symmetry points in the BZ of the tetragonal cell are listed in Table S1 in the Supporting Information (SI).

Spin--orbit coupling (SOC) was included by performing fully relativistic calculations with noncollinear spin \cite{SOC-BSE,SOC-DFT}. 
This treatment is essential due to a presence of heavy $f$-shell impurities.

Strongly correlated systems such as the rare-earth doped oxides require inclusion of Hubbard correction in the both DFT and BSE calculations. 
Hubbard $U$ parameters depend on the radii of the localized orbitals used to evaluate the band occupation numbers, and 
therefore, they are sensitive to a specific implementation of the LDA+$U$ scheme \cite{LDA+U-1,LDA+U-2}. 
For the bulk t-ZrO$_2$, the experimental band gap is known (5.83 eV), thus $U$ values for Zr-$4d$ and O-$2p$ shells can be fitted
to reproduce the band gap; they are 7.2 eV and 3.6 eV, respectively.
Similar parameters were reported by other authors for m-ZrO$_2$: 8 eV for Zr-$4d$ and 4.35 eV for O-$2p$ \cite{U-China,U-calc-ZrO2}. 
Controversely, very different U parameters also fit for the same band gap: 
2.72 eV for Zr-$4d$ and 9.20 or 8.79 eV for O-$2p$ in the first and second neighbor shells, respectively, \cite{U-different}. 
Two sets of U parameters led to the shifts of the valence bands down or conduction bands up, depending on the relative strength of U-Zr-4d and U-O-2p. 
For Eu$^{3+}$ in t-ZrO$_2$, the experimental situation is more complex because the pure crystal without vacancies
does not exist and position of the Eu-$4f$ shell in the band structure is not known. 
Therefore, one can theoretically obtain the Hubbard U from the linear response method \cite{matteo-2005}. 
Hubbard U is defined via trace of a difference between the inverted susceptibility matrices: noninteracting and interacting. 
These matrices are derivatives of the orbital occupation numbers with respect to the perturbation of the potential acting on the Hubbard states.
This method requires using a large supercell of the cell used for the DFT or BSE calculations; this would be not convenient here.
On the other hand, the density functional perturbation theory formulation of the above problem 
enables to use the same cells as for the BSE calculations, and this is
at a price of the derivatives represented in the reciprocal space \cite{hp1,hp2}.
Implementation of this method in {\it hp.x} code \cite{hp} as a part of Quantum ESPRESSO package \cite{QE}
does not allow for freezing the U parameters of some chosen atomic shells (for example Zr-$4d$ and O-$2p$ states).
Hence, using this code, the correction was applied only to Eu in the 'B-Eu' case, 
which led to U=4.35 eV, and this value was used for all bulk calculations. 
Sensitivity of Eu-$4f$ states energy-position with respect to U=3.5 eV and 5 eV is presented in Fig. S1 in SI.
Shift of the deep impurity states in the valence band is estimated for around -1 eV per an increase of the U parameter by 1.5 eV.
As for the pure t-ZrO$_2$ [001] surface, the band gap is almost the same as in the bulk and the same U values were used for Zr-$4d$ and O-$2p$.
The Eu-doped surfaces are too complex for calculations with the hp.x code, not only due to more atoms in the cell but mainly
because of large number of plane waves due to a thick vacuum above the slab.
For this reason, the U-Eu-$4f$ for the surface cases was {\it ad hoc} assumed for 5 eV. 
It is a well known rule that a decrease of the atomic coordination number around impurity enlarges the Hubbard parameter,
usually by 0.5-2 eV, and it is via an increase of the occupation numbers of orbitals not hybridized with the neighboring atoms \cite{UBS1,UBS2,UBS3,UBS4}.
Previous theoretical work about the Eu {\it fcc} metal reported the effective $U$ of $\sim$4--6.5 eV \cite{U-calc-Eu}.

Excitonic properties were calculated by solving the BSE \cite{BSE,MBPT} within
the {\it ab initio} many-body perturbation theory (\textit{ai}-MBPT) framework as implemented in the Yambo code \cite{yambo-1,yambo-2}.
The DFT eigenvalues and eigenvectors were used as input to construct the BSE operators from \textit{first principles}.
The BSE method, in contrast to the DFT, enables to compute and analyze the excited states.
Its MBPT construction, based on the ladder-type Feynman diagrams, includes the electron-hole interaction which is essential for the exciton pair.
For the insulating and semiconductor systems, Tamm-Dancoff approximation \cite{Tamm,Dancoff} makes it possible to transform the BSE,
which is in general the linear-response problem, 
into the eigenvalue problem with a diagonalization of the perturbed Hamiltonian. Hence, the eigenvalues are the exciton energies
and their density of states is the absorption spectrum.  The eigenvectors contain the information on the band-to-band optical transitions
(usually a dipole but a quadrupole also can be chosen) and their amplitudes and weights with k-resolved transitions in BZ.
The huge Hamiltonian matrix can be diagonalized only with the eigenvalues using the Lanczos-Haydock recursion method.
Then, knowing the energetic position of the prominent features, one can perform the diagonalization with vectors  
for up to 20-40 states using the Scalable Library for Eigenvalue Problem Computations  (SLEPc) \cite{slepc}, 
for each of the subsequently chosen 1-2 eV thick parts of the Haydock spectrum. 
This way, the whole energy-axis of the spectrum can be examined for
all locally-strong exciton peaks with the oscillator strength larger than 0.1-0.2 part of the normalized strongest peak.

To solve the BSE, the number of orbitals included in the polarization function exceeded twice the number of electrons in the system.
Plane waves up to 12 Ry were used for the exchange part of the BSE kernel and 4 Ry for the screening and response blocks,
corresponding to up to 4000 RL (reciprocal vectors) for bulk and 16000 RL for surfaces 
(both numbers in RL give the same energy onset of the FFT grid, 12 Ry, due to a large vacuum above the slab).
The $k$-point sampling of the BZ was the same as in the DFT calculations.
Convergence tests of the absorption spectra with respect to the BZ-sampling density and plane-wave cutoffs are presented in Fig.~S2 in SI.
Rare-earth doped systems are less sensitive to BZ sampling than most periodic systems, 
plausibly due to a weak interaction between impurities from the periodic repetition. 
On the other hand, they require a large electron--hole (e--h) interaction space for the accurate BSE calculations.
The number of $k$-points in the irreducible wedge of the BZ (IBZ)
and the number of bands included in the e--h space for all calculated cases are summarized in Table S2 in SI.
The convergence parameters used for the DFT and {\it ai}-BSE calculations are comparable to, or more demanding than,
those employed in the previous theoretical studies of doped ZrO$_2$ \cite{theor-exp-geom,Myrta-ZrO2,Fe-ZrO2}.
The electric-field polarization was chosen along the [1,0,0] crystal direction lying in the basal plane of the tetragonal cell.

Exciton localization in the reciprocal space was analyzed using the Python post-processing Yambopy package \cite{yambopy}. 
Atomic structures as well as electron and hole densities of the selected excitons were visualized in the real space using XCrysDen \cite{xcrysden}.

As for the starting geometry, we note that incorporation of Y and Eu increases the lattice volume \cite{1994,nano-1}. 
Therefore, the experimental lattice constants $a$ = 3.64 \AA\ and $c$ = 5.27 \AA\ \cite{YZO} were used for the doped systems, 
and the atomic positions were relaxed while keeping the cell volume fixed.

\section{Results}

Light absorption spectra obtained with the {\it ai}-BSE for bulk and surface cases are shown in Fig.~4. 
A scheme of the band structure, including the $f$-type states in the valence band (with its width V$_2$--V$_1$), the impurity states 
in the fundamental band gap of the host and the conduction band (manifolds A$_{1,2}$, B$_{1,2}$, and C$_{1,2}$), is presented in Fig.~4(a). 
The $f$-type states can be identified in each case by a projection onto the maximally localized Wannier functions (MLWFs) centered at Eu. 
Projected band structures for all cases with Eu$^{3+}$ and ferromagnetic coupling (for cases with two impurities) 
are presented in Figs.~S3--S8 in SI.
Bands with the $f$-state contribution exceeding 30\% were identified, and their estimated edge energies are listed in Table~S3 in SI. 
The $f-f$ optical transitions correspond to energies $e_j-e_i$, as defined in Fig.~4(a), and are reported for each system and pair of 
$f$-type bands in Table~S3. Realistic $f-f$ optical transitions are expected to be red-shifted relative to the band-to-band transitions by 
the exciton electron--hole binding energy, typically less than 1~eV even for Frenkel excitons. 
Therefore, plausible $f-f$ excitons would appear in the calculated spectra (Figs.~4b--4d) just below the shaded regions, 
shown in violet, orange, and green for the V$\rightarrow$A, V$\rightarrow$B, and V$\rightarrow$C transitions, 
respectively, which represent the energy ranges given in Table~S3.
For all bulk cases, the $f$-states lie deep in the valence band, with the shallowest $f$-states observed in 'S-Eu-V'. 
Optical transitions totally localized at Eu$^{3+}$ are therefore unlikely to occur below 3~eV. 
Excitations from the deep $f$-states would require an extremely intense light source, 
as the electrons are initially excited from the top of the valence band
and a larger work function is necessary to excite electrons from the deep valence states which are strongly hybridized with the neighboring atoms.

\begin{figure*}
        \centerline{\includegraphics[scale=0.25,angle=0.0]{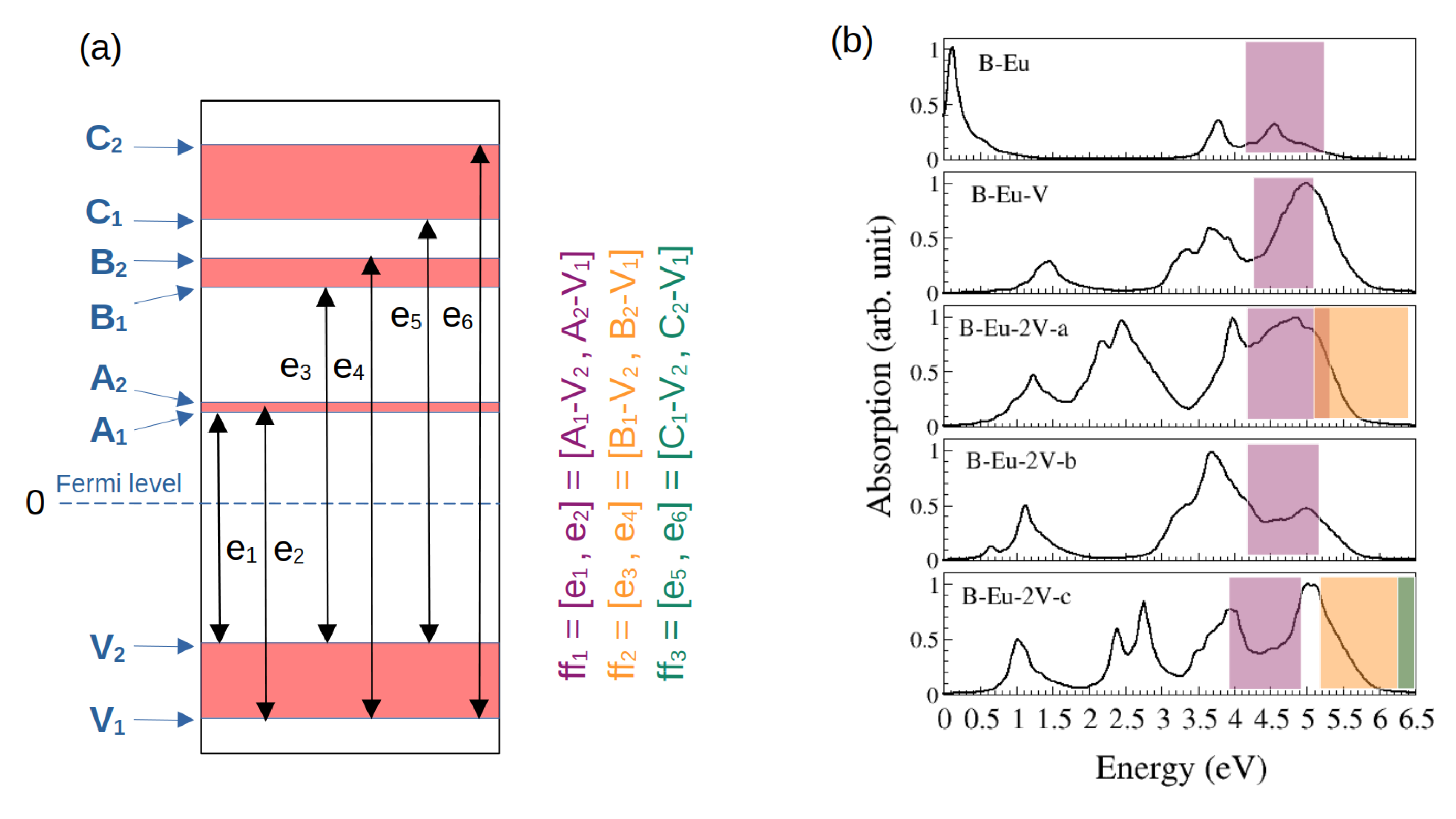}}
        \centerline{\includegraphics[scale=0.25,angle=0.0]{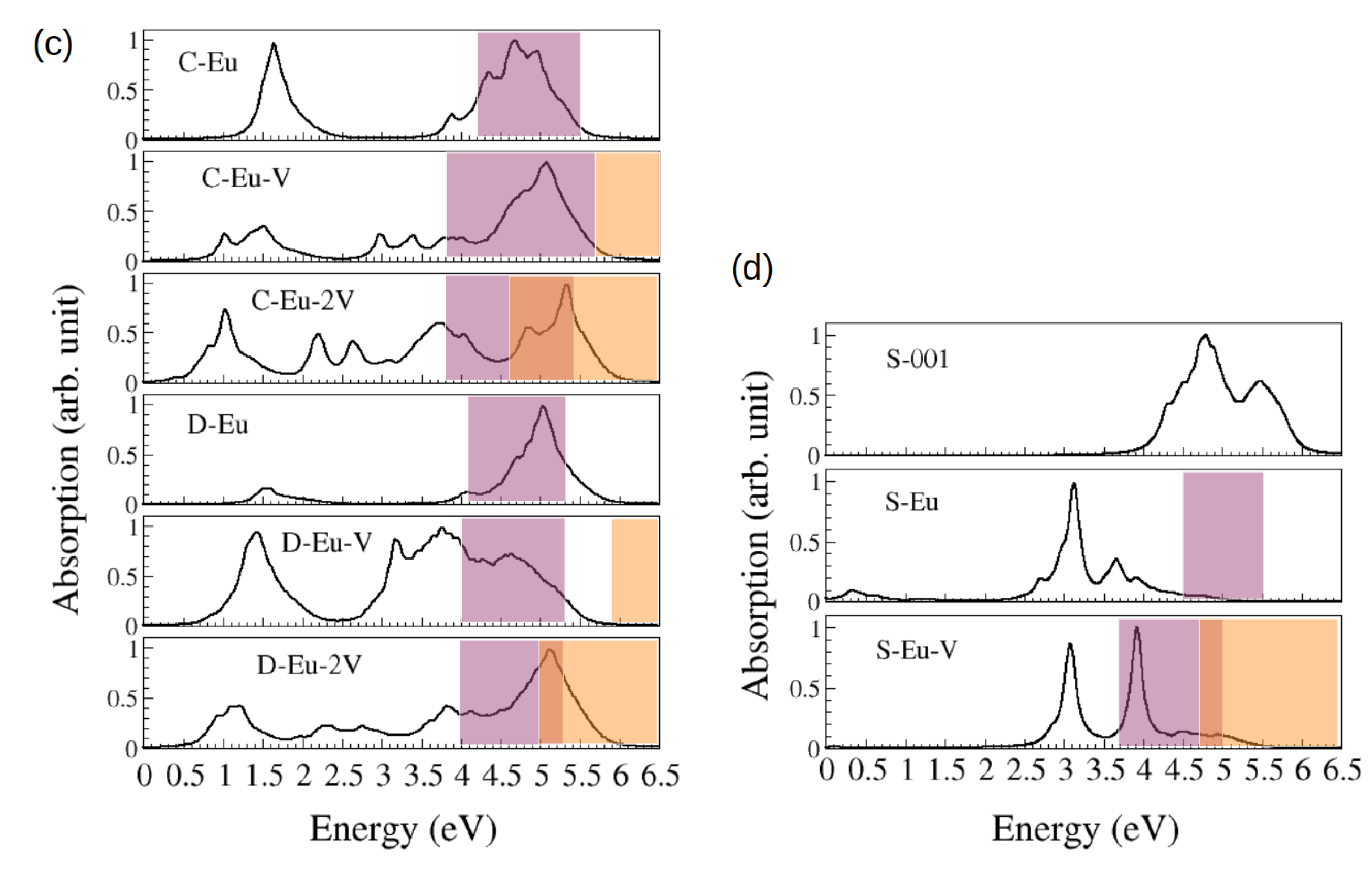}}
	\caption{Mechanism of plausible $f-f$ optical transitions. 
	(a) Scheme of band-to-band transitions between the Eu-$4f$ states: 
	(V$_1$,V$_2$) deep band hybridized with valence states, (A$_1$,A$_2$) pure impurity band within the fundamental band gap,
	(B$_1$,B$_2$) and (C$_1$,C$_2$) impurity states hybridized with conduction states. 
	Absorption spectra calculated with the $ai$-BSE for t-ZrO$_2$: (b) bulk cases with single Eu$^{3+}$,
	(c) bulk cases with two Eu$^{3+}$, (d) [001] surface cases.
	The shaded regions correspond to energies between the Eu-$4f$ occupied and empty states including their bandwidth.}
\end{figure*}

\begin{figure*}
        \centerline{\includegraphics[scale=0.68,angle=0.0]{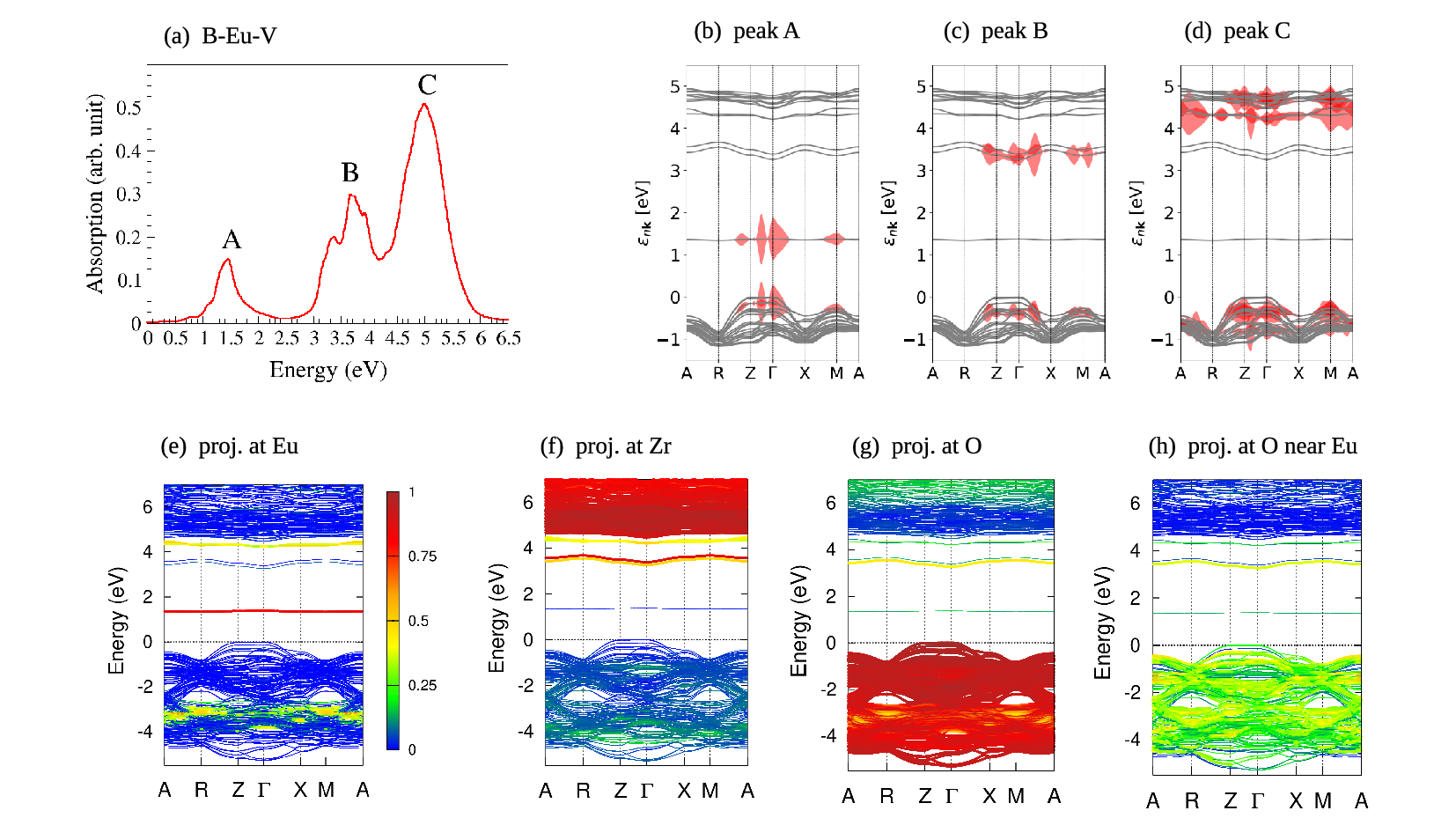}}
        \caption{\label{fig5} Bulk t-ZrO$_2$ with single Eu$^{3+}$ and one V$_\mathrm{O}$ ('B-Eu-V'): 
	(a) absorption spectrum and its three main peaks, 
	(b)-(d) k-resolved transitions of representative excitons in the main peaks, 
	(e)-(h) band structure projected at the MLWF centered at: Eu, all Zr, all O, O near Eu, respectively.
	Color legend is common for all projected band structures.}
\end{figure*}

Further analysis provides insight into exciton spatial localization. As an example, the 'B-Eu-V' case is illustrated in Fig.~5. 
The absorption spectrum, showing three main peaks labeled A, B, and C, is presented in Fig.~5(a). 
Each peak consists of multiple dipole-allowed optical transitions. Figs.~5(b)--5(d) visualize the exciton localization in reciprocal space 
for selected excitons in the main peaks. 
The corresponding band structures are restricted to the chosen electron--hole space included in the {\it ai}-BSE calculations; 
as listed for each case in Table~S2 in SI.
All transitions with the oscillator strength exceeding 50\% of the local maximum are visualized on the corresponding band structures  
in Figs.~S9--S28 in SI alongside the corresponding absorption spectra. Each Eu$^{3+}$ ion in the unit cell introduces one impurity state, 
and each V$_\mathrm{O}$ introduces two impurity states within the fundamental band gap of the host crystal. 
Optical transitions from the valence band into these states occur below $\sim$4.5~eV in the absorption spectrum, 
whereas higher-energy excitations involve transitions between the valence bands and conduction bands of the host.
Regarding exciton localization in BZ, only transitions to the first impurity state exhibit some association 
with the high-symmetry points or lines, albeit with some spread in BZ. 
Higher-energy excitations are highly delocalized in BZ, which is characteristic for Wannier excitons rather than Frenkel excitons. 
To analyze the character of bands involved in the optical transitions, projections onto MLWFs centered at Eu, Zr, all O, 
and selected O atoms near Eu are presented in Figs.~5(e)--5(h). 
In 'B-Eu-V' case, the conduction-band minimum possesses some contribution of Eu states, 
while the valence band is dominated by O states with minor contributions from Zr states (below 25\%).
The Eu states locate in the valence band below $-3$~eV. 
Two V$_\mathrm{O}$ states in the bandgap, near 3.5~eV, are strongly hybridized with Zr. 
Conduction-band manifold primarily consists of Zr states. 
Consequently, the $f-f$ optical transitions are not plausible, due to the lack of Eu states in the shallow valence bands.  
A similar analysis for all cases in Figs.~S8--S27 confirms that the exciton hole is localized on oxygen atoms 
and its electron is localized on Eu and Zr.

Summarizing the absorption peaks in all cases, 
the resulting absorption peaks depend on the number and symmetry of the defect sites,
although all defects in this work correspond to the same charge +2. 
In contrast, other authors have assigned the luminescence features to vacancies 
with multiple charge states: +2, +1, 0, $-1$, and $-2$ \cite{O-blue}. 
Both conclusions highlight a complexity of the experimental reality and suggest that the validation of models solely based on agreement 
between theory and experiment can be misleading. 
This is particularly important when the absorption band is broad and smooth while the characteristic impurity features are minor, 
or when the sample morphology varies \cite{from-Witek,morphology}.

\begin{figure*}
        \centerline{\includegraphics[scale=0.4,angle=0.0]{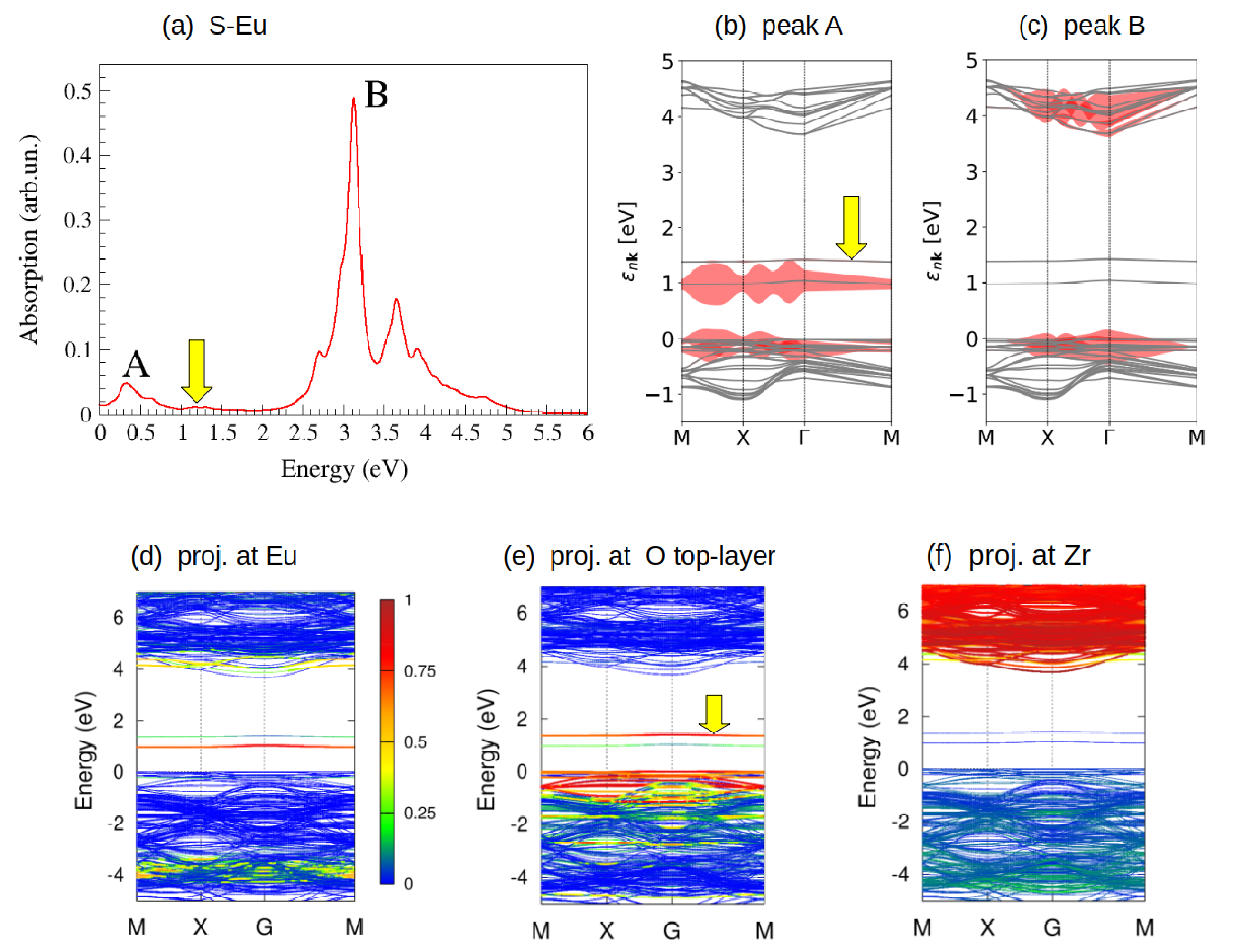}}
	\caption{\label{fig6} The t-ZrO$_2$ [001] surface with single Eu$^{3+}$ ('S-Eu'): 
	(a) absorption spectrum and its two lower-energy peaks,
        (b)-(c)  k-resolved transitions of representative excitons in the corresponding peaks, 
	(d)-(f) band structure projected at MLWF centered at atomic types: Eu, O in top layer, all Zr, respectively.
	Color legend is common for all projected band structures.
	Yellow arrows point to: inactive oxygen-oxygen transitions in absorption spectrum
	and oxygen empty-states in the band structures.} 
\end{figure*}

\begin{figure*}
        \centerline{\includegraphics[scale=0.35,angle=0.0]{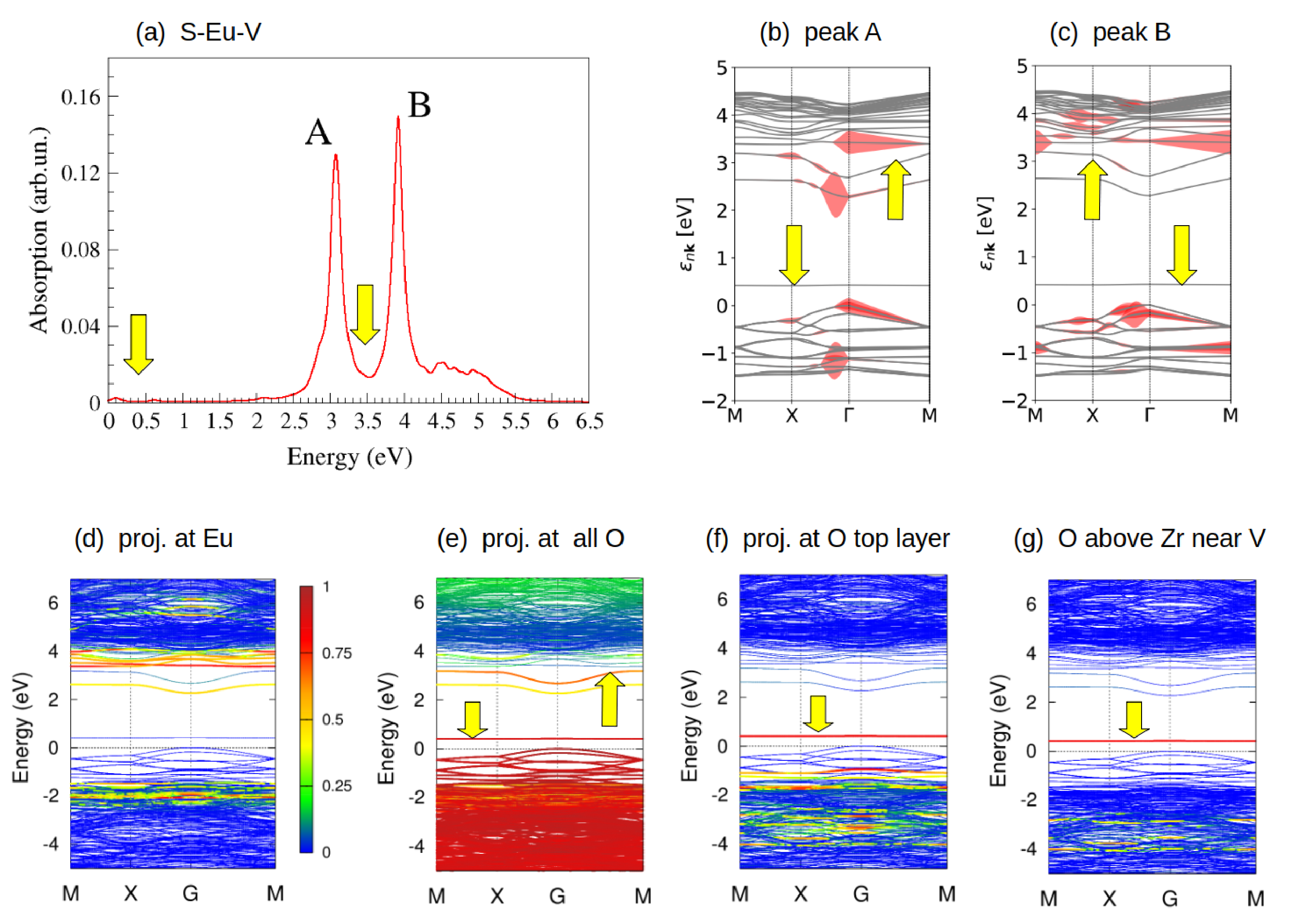}}
	\caption{\label{fig7} The t-ZrO$_2$ [001] surface with single Eu$^{3+}$ and one V$_\mathrm{O}$ ('S-Eu-V'):
        (a) absorption spectrum and its two main peaks,
        (b)-(c) k-resolved transitions of representative excitons in the corresponding peaks,
        (d)-(g) band structure projected at MLWF centered at atomic types: Eu, all O, O in top layer,
        O above Zr near Eu, respectively. Color legend is common for all projected band structures.
        Yellow arrows point to: inactive oxygen-oxygen transitions in absorption spectrum
        and oxygen empty-states in the band structures.}
\end{figure*}

Let us move to the absent peaks which would be expected from the inter-band energies but they do not appear in the spectra.
Probability of the optical transition is mathematically proportional to the integrand of a product of three quantities, 
the dipole (or quadrupole) operator and initial and final state wavefunctions. The symmetry of this these objects leads to a sum
of positive or positive and negative contributions from the real space points, which sometimes may cancell.  
Additionally, some optical transitions to the impurity states within the bandgap seem to be forbidden due to their chemical character,
which is the same as that of the valence bands (i.e. of the oxygen origin).
The empty states of V$_\mathrm{O}$ energetically locate just above, or partially hybridize with, the Eu$^{3+}$ empty states;
see Figs.~S3--S5 in SI. For the hybridized empty states (in the bulk cases), we observe the spectral peaks.
A different character is observed for the surface cases. The 'S-Eu' case exhibits the valence-band maximum 
and conduction-band minimum composed of the top-layer oxygen and Zr states, respectively, 
while the second impurity band (counted from the Fermi level) is of the V$_\mathrm{O}$ origin; see Fig.~S5. 
For 'S-Eu-V', the first and third impurity bands are clearly of the V$_\mathrm{O}$ character: 
the first corresponds to surface oxygens (above Zr near Eu), 
and the third involves oxygens farther from Eu in the lattice; see Fig.~S6.
Consequently, the absorption spectra of 'S-Eu' and 'S-Eu-V' cases, presented in Figs.~6 and 7, do not exhibit peaks corresponding 
to excitations between the oxygen states in the valence band and V$_\mathrm{O}$ empty bands. 
The positions of 'missing' peaks are indicated by yellow arrows, which also point to 
the corresponding V$_\mathrm{O}$ states in the band structures. 
To complete the surface spectra analysis, the strong excitons for 'S001', 'S-Eu' and 'S-Eu-V' are presented in Figs.~S26--S28 in SI.

Turning to the real-space localization of the exciton hole and electron, 
a visualization of the chosen strong excitons for the surface cases is illustrated in Fig.~8.
Panels (a) and (b) show the spatial distribution of electrons for the excitons A and B , respectively, 
in the spectrum of 'S-Eu-V' (from Fig.~7a). 
Excitations in peak A localize their electrons near V$_\mathrm{O}$, whereas peak B involves electrons primarily 
at Eu and partially at Zr. 
Panels (c) and (d) show the corresponding distributions for holes of the same excitons. 
For a comparison, panels (e) and (f) show the hole distributions for peaks A and B of 'S-Eu' (from Fig.~6a).
As a general chemical trend, 
electrons are excited from the regions rich in the negative charge to regions with the reduced electron density. 
For 'S-Eu-V', the first and higher absorption peaks involve excitations from the oxygens which are distant from Eu.
In contrast, for the doped surface without vacancies ('S-Eu'), 
the oxygens which are close to the impurity contribute their electrons to the both low- and high-energy excitation peaks.

\begin{figure}
\includegraphics[scale=0.45,angle=0.0]{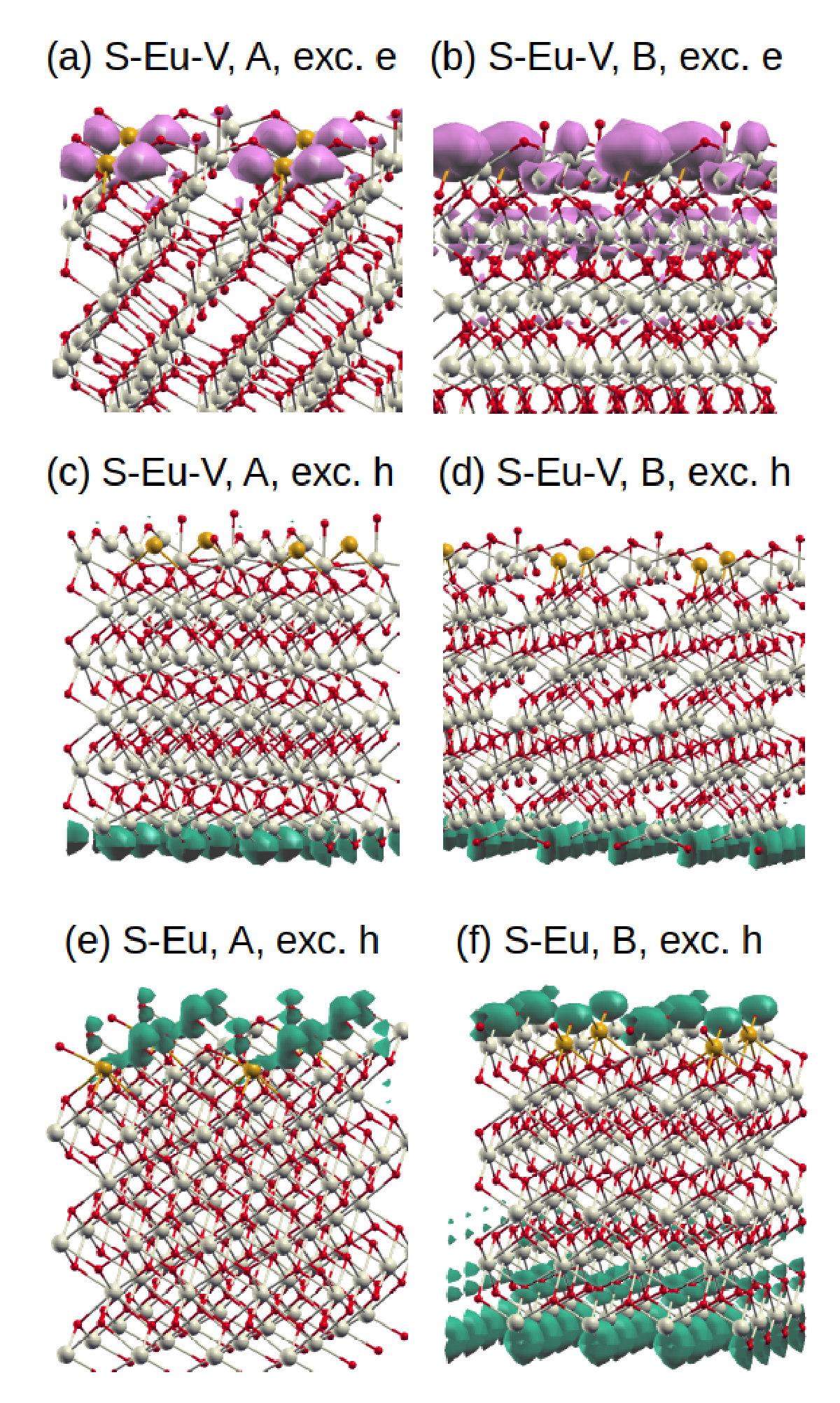}    
        \caption{\label{fig8} Real-space distribution of the electron and hole of the exciton pair: 
	(a)-(d) 'S-Eu-V' case, (e)-(f) 'S-Eu' case, (a)-(b) electron density, (c)-(f) hole density. 
	First peak of spectra, A, is represented in (a), (c) and (e), while the second peak, B, is in (b), (d) and (f).}
\end{figure}

\begin{figure*}
        \centerline{\includegraphics[scale=0.26,angle=0.0]{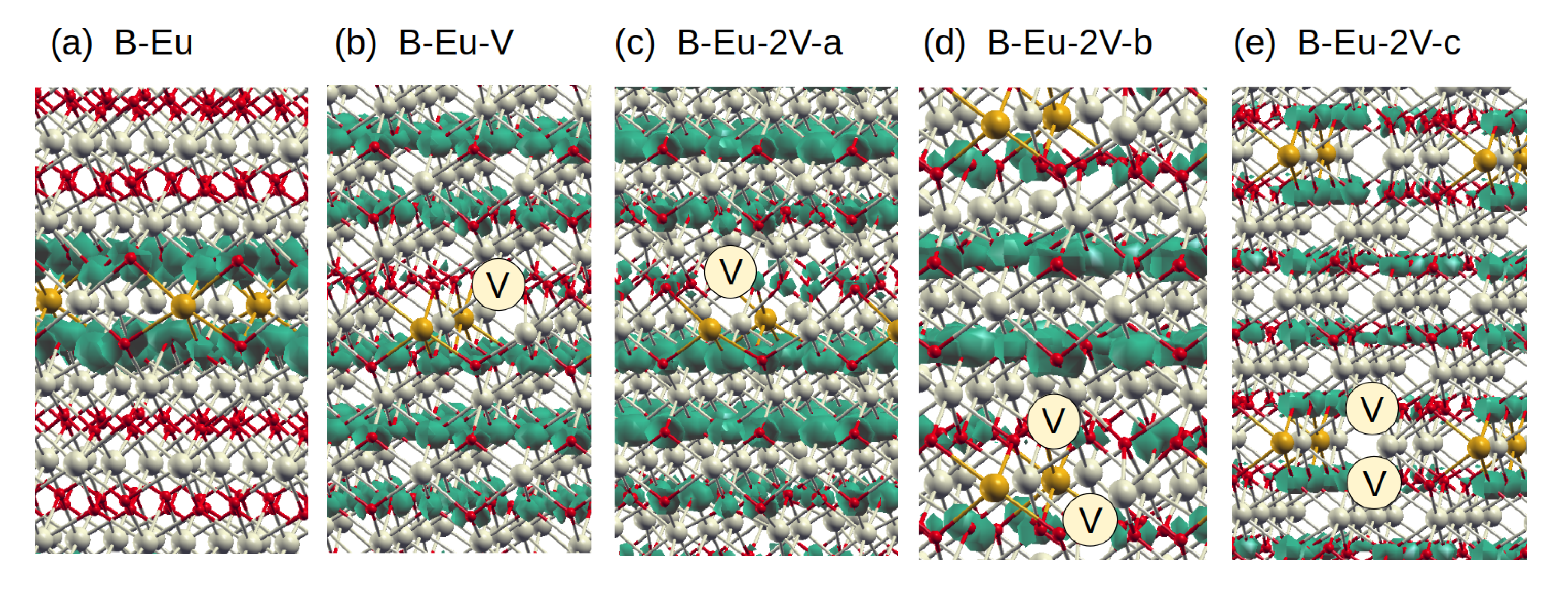}}
        \centerline{\includegraphics[scale=0.26,angle=0.0]{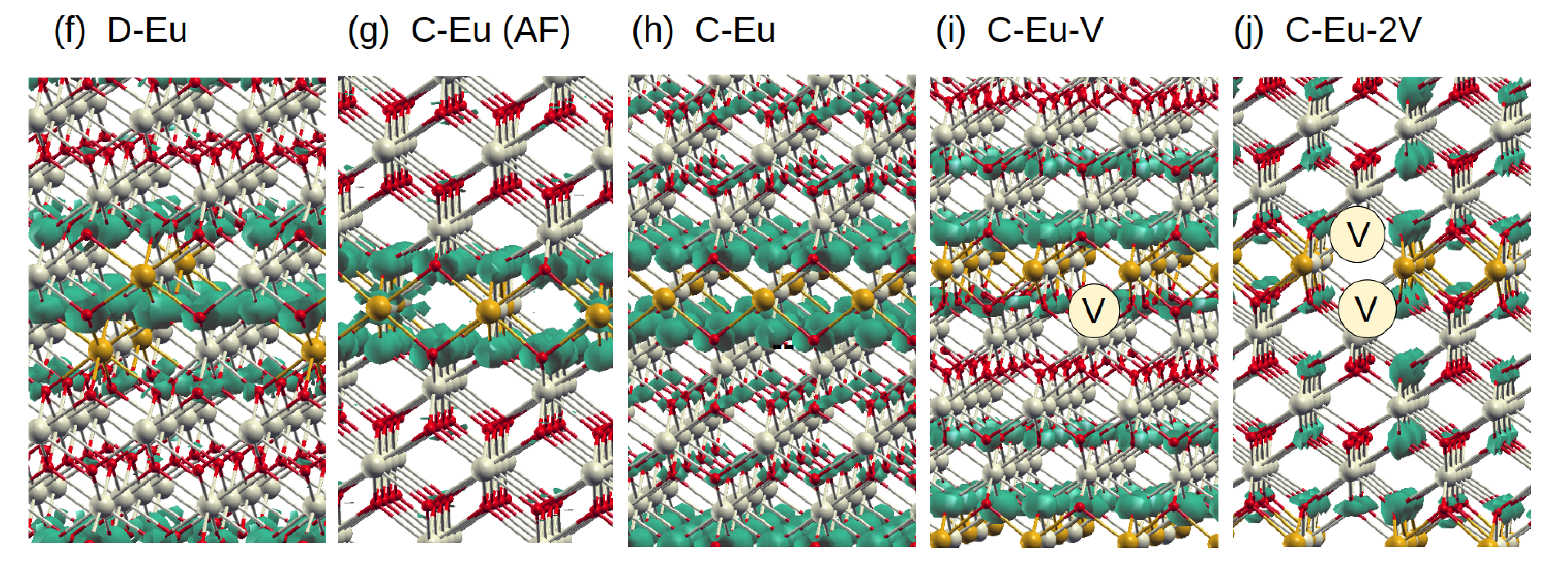}}    
        \caption{\label{fig9} Real-space distribution of a hole of the lowest-energy exciton of bulk t-ZrO$_2$ with:
	(a) single Eu$^{3+}$ ion, (b) single Eu$^{3+}$ and one V$_\mathrm{O}$, 
	(c)-(e) single Eu$^{3+}$ and two V$_\mathrm{O}$ shifted from Eu along axes: 
	$xz$ and $yz$, $xz$ and $yz$, and $xz$ and $yz$, respectively,
	(f) two Eu$^{3+}$ ions in the neighboring Zr-layers, (g)-(j) two Eu$^{3+}$ in the same Zr-layer,
	(i) one V$_\mathrm{O}$, (j) two V$_\mathrm{O}$. 
	All cases with two Eu$^{3+}$ are ferromagnetic, except the (g) case which is antiferromagnetic.} 
\end{figure*}

Figure~9 presents distributions of the holes of the first low-energy excitons in the bulk cases. 
Panels (a)--(e) show cases with a single Eu$^{3+}$ ion, while panels (f)--(j) illustrate selected cases with two Eu$^{3+}$ ions per cell. 
The positions of V$_\mathrm{O}$ (indicated in the corresponding panels) coincide with the regions where the exciton hole density is reduced or absent. 
In cases without V$_\mathrm{O}$, the exciton holes predominantly localize on the oxygen layers adjacent to the layers containing Eu$^{3+}$ ions. 
Thus, the presence of V$_\mathrm{O}$ reduces the exciton localization in bulk systems.

The analysis of light absorption in Eu-doped bulk systems is complemented by the case of a less positively charged ion, Eu$^{2+}$ ('B-Eu-e'); 
see Fig.~S29 in SI. In this case, the hybridization of Eu states with the host is absent, in contrast to the Eu$^{3+}$ ion. 
Seven flat bands appear just below the Fermi level, 
while the host valence band, starting about 2 eV below the impurity states, shows no Eu contribution.  
Weak optical transitions occur between Eu and Zr states up to $\sim$1.5~eV, whereas stronger transitions are above 3~eV 
and involve only the excitations within the host lattice, {\it i.e.} between oxygen and Zr.

Spin effects are significant in the studied systems. Two magnetic couplings, ferromagnetic (FM) and antiferromagnetic (AF), 
between two Eu$^{3+}$ ions in the bulk cases were compared; see Table~S4 and Fig.~S30 in SI. 
Magnetization in doped systems is generally localized at 
the impurities and characterized by the total ($M_\mathrm{tot}$) and absolute ($M_\mathrm{abs}$) magnetizations, defined via ithe spin-polarized densities. 
In the calculations with SOC, the diagonal elements of the density matrix, $\rho_{11}({\bf r})$ and $\rho_{22}({\bf r})$, are expressed 
in terms of Pauli matrices $[{\bf \sigma}_\alpha]_{ss'}$ and spinors $\phi_{n{\bf k},s}$ for the occupied bands $n$ and all $k$-points in BZ. 
A set of the following equations
\begin{eqnarray}
M_\mathrm{tot} & = & \int \rho_{11}({\bf r}) \, d{\bf r} - \int \rho_{22}({\bf r}) \, d{\bf r}, \\
M_\mathrm{abs} & = &  \int \left| \rho_{11}({\bf r}) - \rho_{22}({\bf r}) \right| \, d{\bf r}, \\
\rho_{ss'}({\bf r}) & = & \sum_{n{\bf k}} \sum_{\alpha=0,3} \phi_{n{\bf k},s}^{\ast}({\bf r}) \, [{\bf \sigma}_{\alpha}]_{ss'} \, \phi_{n{\bf k},s'}({\bf r}), \quad \{s,s'\} \in \{1,2\}.
\end{eqnarray}
has been used for all studied cases; presented in Table S4 in SI.
Noncollinear magnetization due to the SOC introduces the substantial $x$ and $y$ components of $M_\mathrm{tot}$ in cases with two V$_\mathrm{O}$ per cell. 
The local magnetization at Eu$^{3+}$ is $\sim$6~$\mu_B$ in bulk, $\sim$5~$\mu_B$ at the surface without vacancies, 
and $\sim$6~$\mu_B$ at the surface with vacancies, whereas Eu$^{2+}$ in bulk has a spin moment of $\sim$7~$\mu_B$. 
Values of $M_\mathrm{abs}$ exceeding those expected from the number and coupling of Eu$^{3+}$ ions reflect 
the induced magnetization of neighboring atoms around V$_\mathrm{O}$.
Turning to the absorption spectra of bulk systems with two Eu$^{3+}$ ions per cell, presented in Fig.~S30 SI, 
substancial differences between FM and AF couplings are observed. 
However, since the electron-hole space was restricted to 1.5~eV below and above the valence and conduction bands, 
the part of spectra above the first host-to-host exciton peak ($\sim$4.5--5.0~eV) should not be compared directly. 
For 'D-Eu-V', the first exciton peak (a charge-transfer exciton O-2p$\rightarrow$Eu) 
is lower for AF than FM coupling, whereas this trend reverses when both Eu$^{3+}$ ions are in the same layer ('C-Eu-V'). 
This behavior accidentally relates to the SOC-induced spin mixing, 
which is analyzed for all bulk cases in Figs.~S31--S33 in SI (band structures projected onto $S_z$). 
FM cases exhibit nearly pure spin character, whereas AF cases show the strong spin mixing for 'C-Eu-V' and 'D-Eu-2V',  
a moderate spin mixing for 'C-Eu-2V', and weak spin mixing for 'C-Eu', 'D-Eu', and 'D-Eu-V'. 
A 'phenomenological' observation is that the strong spin mixing shifts the absorption peaks associated with the impurity excitations. 
This fact would be consistent with an intuition that both spin channels of the electron-donating oxygen atoms are active 
within the restricted energy range of the topmost valence bands.
Importantly, the reliable conclusions within a rigorous mathematical analysis of excitons cannot be performed without a tool that combines the exciton
dipole transitions weights with the spin projections of the involved bands and k-points in BZ.

As for a comparison of the calculated spectra with the experiment, it is difficult to find reports on direct measurement of light absorption,
except a very narrow-IR-range plotted for phonon absorption \cite{Lojek}. There was also a short note on transient absorption in ZrO$_2$:Y
(sec 4.4 in PhD thesis \cite{SmitsPhD}). Author concluded about two transient absorption bands at 1.7 eV and 2.6 eV.
Comparison of the luminescence decay under electron beam excitation (in nanoseconds) and transient absorption decay in time (long lasting) led
in that dissertation to the conclusion that the centers responsible for luminescence and transient absorption are not the same.
This fact might be important for an interpretation of the luminescence excitation spectra because their direct comparison with absorption 
is not reliable. The rare-earth doped systems possess many optical transition that relax also via vibrations (phonon), thus the excitation spectra
for two emission lines, even energetically close, very often substantially differ one from the other for some peaks.
The absorption edge of ZrO$_2$ doped with Eu$^{3+}$ at 0.4\%, 1\%, 2\% and 4\% was estimated from the reflection spectra \cite{chemisensors} (Fig. 4 therein)
pointing to an increased absorption intensity and slight blue-shift for the growing doping rates. These data were sensitive only to the charge-transfer
optical transitions O-2p$\rightarrow$Eu$^{3+}$-4f (around 4.13 eV and higher energies). Additionally, the above work presented the excitation spectra
down to 500 nm (2.48 eV) that contained some low-intensity peaks (Fig. 7 therein) attributed by the authors to at least two types of Eu ions at different
oxidation states. 
The luminescence excitation spectra were reported also in Refs. \cite{exp-Borik} and \cite{exp-jalcom}.
Authors of the former work examined two samples in the tetragonal phase with some monoclinic contribution, at temperature 300 K, for two emission
lines: 606 nm and 612 nm for the sample 1 and 425 nm and 431.8 nm for the sample 2. They concluded about signatures of Eu$^{3+}$ in the both crystal
phases for the sample 1 and existence of the lines characteristic for Eu$^{2+}$ and weak contamination of Eu$^{3+}$ for the sample 2 (see Figs. 6 and 12
in Ref. \cite{exp-Borik}). 
Authors of the latter work described samples of ZrO$_2$:Eu$^{3+}$ in different morphology, nanotubes and nanopowder, in the tetragonal and monoclinic phases. 
They reported the constant energetic position of the corresponding peaks measured in all samples (for all morphology) 
and that only the peak intensities varied (see Fig. 4 in Ref. \cite{exp-jalcom}).

One can make inferences from the experimental and JOT-based results, that authors assign all detected peaks according to the free-ion
reference tables adding the crystal-field shifts to the peaks. Plausibly, experiments register mainly luminescence of
the surface ions, while the impurities embedded in the bulk (similar to cases reported in this work) give the 
less intense signals, due to their statistical character, and sum to the averaged background in the measured spectra. 
For a future work, the BSE spectra calculated for the surfaces other than [001] would bring the valuable additional information.

\section{Conclusion}

Tetragonal ZrO$_2$ doped with Eu$^{3+}$ in the bulk and on the [001] surface, both with and without single and double oxygen vacancies 
in various symmetry configurations, have been studied using the {\it ai}-BSE including Hubbard $U$ correction and spin--orbit coupling. 
Light absorption spectra were analyzed to assess a plausibility of the $f-f$ excitations localized at individual, or between two, Eu ions.

The results allow us to formulate the following rules: 
Holes of the exciton pairs are distributed over oxygen atoms originating from the negative-charge rich regions of the crystal, 
{\it i.e.} parts with fewer oxygen vacancies. 
Electrons of the exciton pairs are localized at Eu for the low-energy peaks and mainly at Zr (with some contribution from Eu where such states exist) 
for optical transitions across the fundamental band gap. 
The Eu-$4f$ states hybridized with the valence band are at energies below $-3$ eV in the bulk cases 
and $-1.6$ eV for the [001] surface with one oxygen vacancy per Eu. 
Therefore, the impurity states lie deep enough in the valence band to be most probably inactive for the optical excitations
in the IR and visible light range; although the accurate band structures depend on a choice of the Hubbard $U$.
Figure 10 summarizes positions of the Eu-$4f$ states in the valence band and the character
of dipole transitions in the absorption spectra.
Concentrations of Eu$^{3+}$ (with respect to all cation sites) are 6$\%$ and 12.5$\%$ for singly and doubly doped bulk cells, respectively. 
The doping level almost does not affect the energetic position of the impurity states, as presented in Fig. S34 in SI
for a doubly-doped supercell enlarged by undoped cell of the same size. However, a hybridization of the impurity states with the host is 
stronger when impurities are separated on a larger distance. In turn, this strong hybridization with the valence bands require a larger number
of photons in the volume to be efficient for the impurity excitation. 
\begin{figure*}
        \centerline{\includegraphics[scale=0.22,angle=0.0]{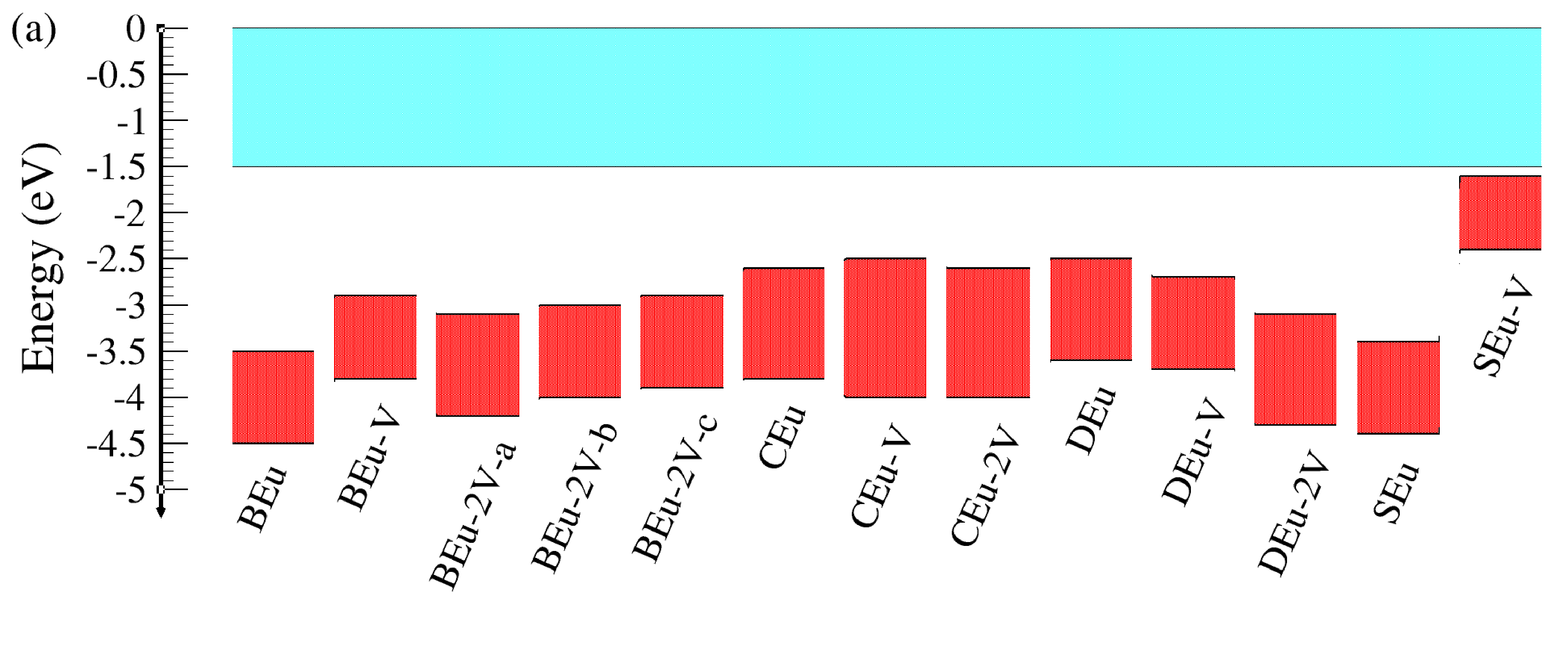}}
        \centerline{\includegraphics[scale=0.3,angle=0.0]{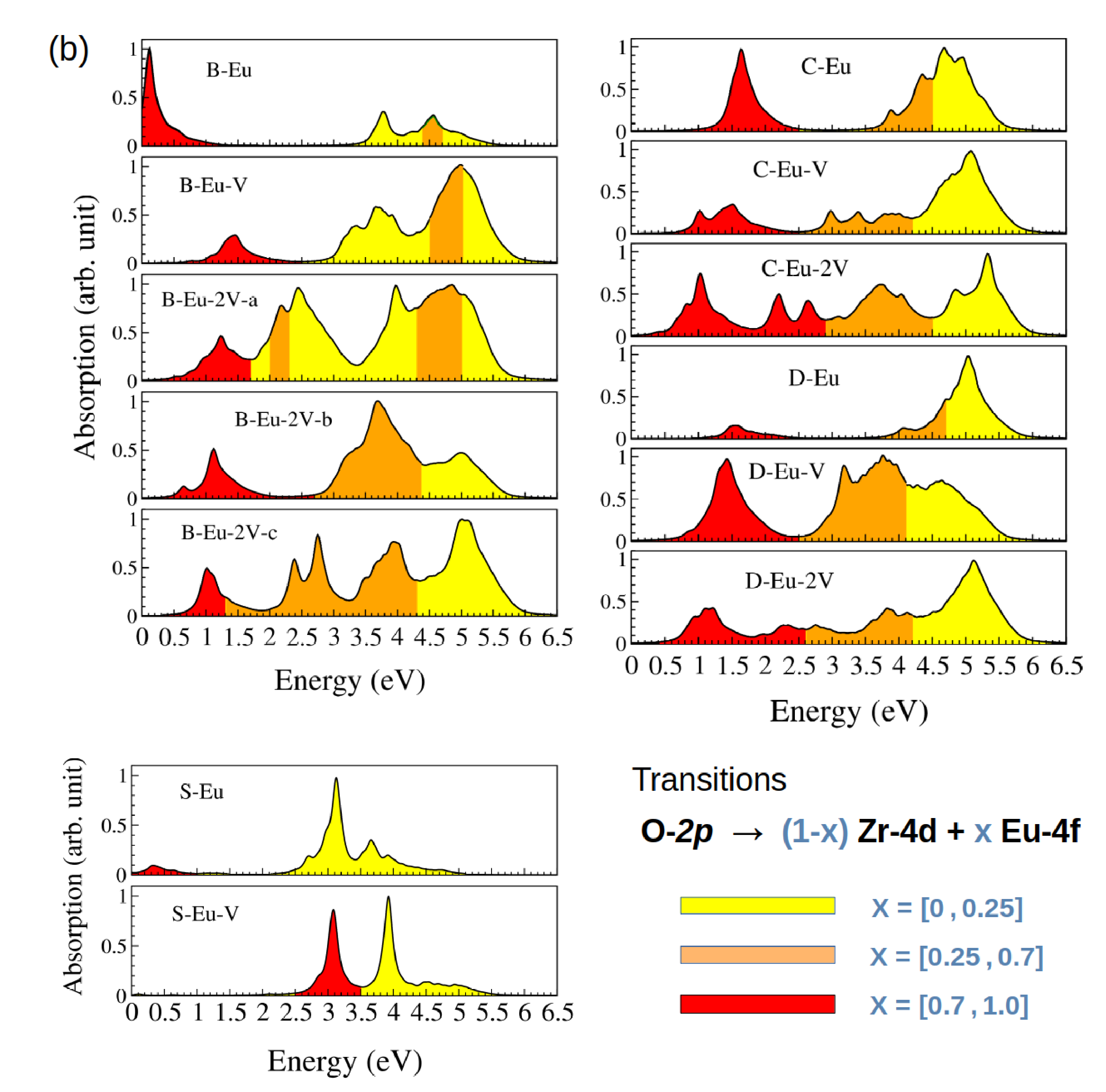}}
        \caption{\label{fig10} (a) Energy position and thickness of Eu-$4f$ states hybridized with the valence band 
        (red color) and valence band range included in the BSE e--h space (cyan color).
        (b) Absorption spectra analyzed for a character of dipole transitions, according to the legend.}
\end{figure*}

Both ferromagnetic (FM) and antiferromagnetic (AF) couplings between two Eu$^{3+}$ ions in the unit cell were considered. 
A 'numerical' observation is that
the AF order is more effective than FM for light absorption in the cases with strong SOC. 
This fact numerically correlates with the $S_z$-component mixing in the spinors 
and the significant planar contribution to the total magnetization vector observed for some cases with oxygen vacancies.

Additionally, for Eu$^{2+}$ in the bulk, the low-energy excitations in absorption are weak and occur between Eu-$4f$ states in the valence band 
and Zr-$4d$ states in the conduction band. This is consistent with the fact that the Eu$^{2+}$ ion replaces Zr$^{4+}$ 
and introduces additional negative charge acting as a donor for the optical transitions.

{\it Ab initio} calculations reveal a scenario that definitively excludes the pure Eu-$4f$ localized optical transitions. 
This contrasts with the phenomenological studies based on the Judd--Ofelt model reported for rare-earth doped systems
\cite{tZrO2-orange,model-1,nano-2,Brik-model,nano-1,from-Witek,morphology,lumin-ff-ZrO2,O-blue,oxide-2,glass-1,glass-2,glass-3,glass-2024}. 
The fundamental difference between the {\it ai}-MBPT and Judd--Ofelt approach is lying in the assumed origin of the excited electrons: 
the valence-band top in the former and impurity states in the latter. 
Experimental temperature-dependent absorption and emission spectra of crystalls and glasses generally support a charge-transfer scenario
\cite{temp-1,temp-2,temp-3,temp-4},
because strong temperature effects are closely related to phonons and these, in turn, operate solely between and among atoms and not on-site. 

\begin{acknowledgments}
I acknowledge EPICURE, the EuroHPC Joint Undertaking initiative, for supporting the scalability tests of the codes used in this project 
on the LUMI supercomputer through the EuroHPC Benchmark Access call (grant EHPC-BEN-2025B02-048). 
Calculations were also performed under PL-Grid grants on LUMI (PLL/2024/07/017622) and Ares (PLG/2024/017753).
\end{acknowledgments}

\bibliography{lib.bib}  

@article{glass-1,
  title = {Comparative study on the influence of rare earth ion doping on the structural and optical properties of simple {B$_2$O$_3\--$Na$_2$O} glasses},
  volume = {15},
  ISSN = {2045-2322},
  DOI = {10.1038/s41598-025-21224-x},
  number = {1},
  journal = {Scientific Reports},
  publisher = {Springer Science and Business Media LLC},
  author = {El-shabaan,  M. M. and Mohamed,  Amaal and Youssif,  M. I. and El-Ghamaz,  N. A. and Ahmed,  E. M.},
  year = {2025},
  pages = "35022"
}

@article{glass-2,
  title = {Luminescence properties and quantum efficiency of the {Eu}-doped borate glasses},
  volume = {77},
  ISSN = {0925-3467},
  DOI = {10.1016/j.optmat.2018.01.019},
  journal = {Optical Materials},
  publisher = {Elsevier BV},
  author = {Kindrat,  I.I. and Padlyak,  B.V.},
  year = {2018},
  month = mar,
  pages = {93--103}
}

@article{glass-3,
  title = {Development of {Eu$^{3+}$}-doped phosphate glass for red luminescent solid-state optical devices},
  volume = {227},
  ISSN = {0022-2313},
  DOI = {10.1016/j.jlumin.2020.117564},
  journal = {Journal of Luminescence},
  publisher = {Elsevier BV},
  author = {Aryal,  P. and Kim,  H.J. and Khan,  A. and Saha,  S. and Kang,  S.J. and Kothan,  S. and Yamsuk,  Y. and Kaewkhao,  J.},
  year = {2020},
  month = nov,
  pages = {117564}
}

@article{organic-1,
  title = {Structural and Luminescence Properties of {Eu}-Doped PMO Films with Ethylene Bridge and Methyl Terminal Groups},
  volume = {13},
  ISSN = {2079-6412},
  DOI = {10.3390/coatings13091656},
  number = {9},
  journal = {Coatings},
  publisher = {MDPI AG},
  author = {Rasadujjaman,  Md and Zhang,  Jinming and Vishnevskiy,  Alexey S. and Zhang,  Jing and Baklanov,  Mikhail R.},
  year = {2023},
  month = sep,
  pages = {1656}
}

@article{app-radiation,
  title = {Optical features and radiation protection capabilities of BaO and {Eu$_2$O$_3$} doped borate-based glasses: The effect of composition on performance},
  volume = {53},
  ISSN = {2307-4108},
  DOI = {10.1016/j.kjs.2025.100483},
  number = {1},
  journal = {Kuwait Journal of Science},
  publisher = {Elsevier BV},
  author = {Alrowaily,  Albandari W. and Manjunatha and Biradar,  Shrikant and Sayyed,  M.I. and Bennal,  A.S.},
  year = {2026},
  month = jan,
  pages = {100483}
}

@article{upconversion,
  title = {Upconversion and Color Tunability in {Er$^{3+}$}\--{Tm$^{3+}$}\--{Yb$^{3+}$} Tri-Doped Fluorophosphate Glasses},   
  volume = {12},
  ISSN = {2304-6732},
  DOI = {10.3390/photonics12080745},
  number = {8},
  journal = {Photonics},
  publisher = {MDPI AG},
  author = {Rivera-L/'opez,  Fernando and Babu,  Palamandala and Venkatramu,  Vemula and Lav/'in,  Víctor},
  year = {2025},
  month = jul,
  pages = {745}
}

@article{temp-1,
  title = {Processing and luminescence of {Eu/Dy}-doped {Sr$_2$MgSi$_2$O$_7$} glass-ceramics},
  volume = {41},
  ISSN = {0955-2219},
  DOI = {10.1016/j.jeurceramsoc.2020.08.038},
  number = {1},
  journal = {Journal of the European Ceramic Society},
  publisher = {Elsevier BV},
  author = {Fern/andez-Rodr/'iguez,  L. and Levy,  D. and Zayat,  M. and Jim/'enez,  J. and Mather,  G.C. and Dur/'an,  A. and Pascual,  M.J.},
  year = {2021},
  month = jan,
  pages = {811--822}
}

@article{UV-sensor,
  title = {Quantum efficiency of europium doped {LaPO$_4$} phosphors for UV sensing applications},
  volume = {14},
  ISSN = {2080-2242},
  DOI = {10.4302/plp.v14i2.1146},
  number = {2},
  journal = {Photonics Letters of Poland},
  publisher = {Photonics Society of Poland},
  author = {Sadowska,  Karolina and Awramiuk,  Pawe/l{} and Zg/l{}obicka,  Izabela and Re/'cko,  Katarzyna and /.Zmojda,  Jacek},
  year = {2022},
  month = jul,
  pages = {28}
}

@article{UV-devices,
  title = {Hydrothermal synthesis and vacuum ultraviolet-excited luminescence properties of novel {Dy$^{3+}$}-doped {LaPO$_4$} white light phosphors},
  volume = {44},
  ISSN = {0025-5408},
  DOI = {10.1016/j.materresbull.2009.07.021},
  number = {12},
  journal = {Materials Research Bulletin},
  publisher = {Elsevier BV},
  author = {Han,  Guocai and Wang,  Yuhua and Wu,  Chunfang and Zhang,  Jiachi},
  year = {2009},
  month = dec,
  pages = {2255--2257}
}

@article{envir-probe,
  title = {Luminescent Studies on Germanate Glasses Doped with Europium Ions for Photonic Applications},
  volume = {13},
  ISSN = {1996-1944},
  DOI = {10.3390/ma13122817},
  number = {12},
  journal = {Materials},
  publisher = {MDPI AG},
  author = {/.Zmojda,  Jacek and Kochanowicz,  Marcin and Miluski,  Piotr and Golonko,  Piotr and Baranowska,  Agata and Ragi/'n,  Tomasz and Dorosz,  Jan and Kuwik,  Marta and Pisarski,  Wojciech and Pisarska,  Joanna and Szal,  Renata and Mach,  Gabriela and Starzyk,  Bartosz and Le/'nniak,  Magdalena and Sitarz,  Maciej and Dorosz,  Dominik},
  year = {2020},
  month = jun,
  pages = {2817}
}

@article{opt-temp,
  title = {{Eu$^{3+}$} doped {CaWO$_4$} nanophosphor for high sensitivity optical thermometry},
  volume = {305},
  ISSN = {1386-1425},
  DOI = {10.1016/j.saa.2023.123542},
  journal = {Spectrochimica Acta Part A: Molecular and Biomolecular Spectroscopy},
  publisher = {Elsevier BV},
  author = {Xiao,  Haihong and Meng,  Qingyu},
  year = {2024},
  month = jan,
  pages = {123542}
}

@article{Eu-LED,
  title = {Europium-Doped Calcium Silicate Nanoparticles as High-Quantum-Yield Red-Emitting Phosphors},
  volume = {6},
  ISSN = {2574-0970},
  DOI = {10.1021/acsanm.3c01547},
  number = {11},
  journal = {ACS Applied Nano Materials},    
  publisher = {American Chemical Society (ACS)},     
  author = {Woo,  Hyun-Joo and Chung,  Seock-Jin and Hill,  Meghan L. and Hadrick,  Kay and Kim,  Taeho},     
  year = {2023}, 
  month = may,   
  pages = {9884--9891}     
}

@article{blue-red,
  title = {Tuning of photoluminescence by co-doping {Eu$^{2+}$},  {Eu$^{3+}$} and {Tb$^{3+}$} in {Ca$_9$NaZn(PO$_4$)$_7$} phosphor},
  volume = {150},
  ISSN = {0143-7208},
  DOI = {10.1016/j.dyepig.2017.12.026},
  journal = {Dyes and Pigments},
  publisher = {Elsevier BV},
  author = {Wu,  Zhan-Chao and Wang,  Fang-Fang and Zhang,  Kun and Xu,  Shi-Cheng and Liu,  Jie and Xia,  Zhiguo and Wu,  Ming-Mei},
  year = {2018},
  month = mar,
  pages = {275--283}
}

@article{cryst-struc,
  title = {Photoluminescence properties of {Eu$^{3+}$} doped {ZrO$_2$} with different morphologies and crystal structures},
  volume = {864},
  ISSN = {0925-8388},
  DOI = {10.1016/j.jallcom.2021.158781},
  journal = {Journal of Alloys and Compounds},
  publisher = {Elsevier BV},
  author = {Liu,  Dongxin and Wang,  Mingli and Gong,  Liyuan and Zhao,  Jianling and Zhu,  Miao and Wang,  Xixin},
  year = {2021},
  month = may,
  pages = {158781}
}

@article{tZrO2-orange,
  title = {Combustion synthesized tetragonal {ZrO$_2$}: {Eu$^{3+}$} nanophosphors: Structural and photoluminescence studies},
  volume = {135},
  ISSN = {1386-1425},
  DOI = {10.1016/j.saa.2014.06.151},
  journal = {Spectrochimica Acta Part A: Molecular and Biomolecular Spectroscopy},
  publisher = {Elsevier BV},
  author = {Vidya,  Y.S. and Anantharaju,  K.S. and Nagabhushana,  H. and Sharma,  S.C. and Nagaswarupa,  H.P. and Prashantha,  S.C. and Shivakumara,  C. and Danithkumar},
  year = {2015},
  month = jan,
  pages = {241–251}
}

@article{Eu2BaO,  
  title = {Enhanced photoluminescence from mixed-valence {Eu}-doped nanocrystalline silicate glass ceramics}, 
  volume = {18},   
  ISSN = {1094-4087},   
  DOI = {10.1364/oe.18.00a575},    
  number = {S4},    
  journal = {Optics Express},    
  publisher = {Optica Publishing Group},    
  author = {Gao,  Guojun and Da,  Ning and Reibstein,  Sindy and Wondraczek,  Lothar},     
  year = {2010},   
  month = oct,   
  pages = {A575}   
}

@article{nano-1,
  title = {Shaping the photoluminescence spectrum of {ZrO$_2$:Eu$^{3+}$} phosphor in dependence on the {Eu} concentration},
  volume = {121},
  ISSN = {0925-3467},
  DOI = {10.1016/j.optmat.2021.111620},
  journal = {Optical Materials},
  publisher = {Elsevier BV},
  author = {Nashivochnikov,  Aleksandr A. and Kostyukov,  Anton I. and Zhuzhgov,  Aleksey V. and Rakhmanova,  Marianna I. and Cherepanova,  Svetlana V. and Snytnikov,  Valeriy N.},
  year = {2021},
  month = nov,
  pages = {111620}
}

@article{Sc-Ce-Y,
  title = {Structure and transport properties of zirconia crystals {Co}-doped by scandia,  ceria and yttria},
  volume = {5},
  ISSN = {2352-8478},
  DOI = {10.1016/j.jmat.2019.02.004},
  number = {2},
  journal = {Journal of Materiomics},
  publisher = {Elsevier BV},
  author = {Agarkov,  D.A. and Borik,  M.A. and Bredikhin,  S.I. and Burmistrov,  I.N. and Eliseeva,  G.M. and Kolotygin,  V.A. and Kulebyakin,  A.V. and Kuritsyna,  I.E. and Lomonova,  E.E. and Milovich,  F.O. and Myzina,  V.A. and Ryabochkina,  P.A. and Tabachkova,  N.Yu. and Volkova,  T.V.},
  year = {2019},
  month = jun,
  pages = {273--279}
}

@article{nano-2,
  title = {The synthesis and fluorescent properties of nanoparticulate {ZrO$_2$} doped with {Eu} using continuous hydrothermal synthesis},
  volume = {11},
  ISSN = {1463-9270},
  DOI = {10.1039/b812149d},
  number = {4},
  journal = {Green Chemistry},
  publisher = {Royal Society of Chemistry (RSC)},
  author = {Hobbs,  Helen and Briddon,  Stephen and Lester,  Ed},
  year = {2009},
  pages = {484}
}

@article{Myrta-ZrO2,
  title = {Quasiparticle calculations of the electronic properties of {ZrO$_2$} polymorphs and their interface with {Si}},
  volume = {81},
  ISSN = {1550-235X},
  DOI = {10.1103/physrevb.81.035330},
  number = {3},
  journal = {Physical Review B},
  publisher = {American Physical Society (APS)},
  author = {Gr\"{u}ning,  Myrta and Shaltaf,  Riad and Rignanese,  Gian-Marco},
  year = {2010},
  month = jan 
}

@article{ZrO2-geom,  
  title = {Study on Structural Stability of {ZrO$_2$} and {YSZ}: Doping-Induced Phase  Transitions and Fermi Level Shift}, 
  ISSN = {2717-5804}, 
  DOI = {10.37256/aecm.5120243686},
  journal = {Advanced Energy Conversion Materials},
  publisher = {Universal Wiser Publisher Pte. Ltd},
  author = {D. D. Nematov and A. S. Burkhonzoda and Kh T. Kholmurodov and A. I. Lyubchyk and S. I. Lyubchyk and S. B. Lyubchyk},
  year = {2023},
  month = nov,
  pages = {1--20}
}

@article{oxide-2,
  title = {Photoluminescence Properties and Quantum Efficiency of {Eu$^{3+}$/Mn$^{2+}$}\-- Doped {ZnMoO$_4$} Red Phosphor},
  volume = {843},
  ISSN = {1662-9795},
  DOI = {10.4028/www.scientific.net/kem.843.70},
  journal = {Key Engineering Materials},
  publisher = {Trans Tech Publications,  Ltd.},
  author = {Chen,  Fan and Balci,  Mustafa Hasan and Xia,  He Xin and Akram,  Muhammad Nadeem and Chen,  Xu Yuan},
  year = {2020},
  month = may,
  pages = {70--78}
}

@article{Eu2YAG,
  title = {{Eu$^{2+}$} Stabilization in YAG Structure: Optical and Electron Paramagnetic Resonance Study},
  volume = {120},
  ISSN = {1932-7455},
  DOI = {10.1021/acs.jpcc.6b06397},
  number = {38},
  journal = {The Journal of Physical Chemistry C},
  publisher = {American Chemical Society (ACS)},
  author = {Havl/'ak,  L. and B/'arta,  J. and Buryi,  M. and Jar/'y,  V. and Mih/'okov/'a,  E. and Laguta,  V. and Boh/'a/cek,  P. and Nikl,  M.},
  year = {2016},
  month = sep,
  pages = {21751--21761}
}

@article{model-1,
  title = {Spectroscopic studies of {Eu$^{3+}$} ions in {LBTAF} glasses},
  volume = {478},
  ISSN = {0925-8388},
  DOI = {10.1016/j.jallcom.2008.12.013},
  number = {1–2},
  journal = {Journal of Alloys and Compounds},
  publisher = {Elsevier BV},
  author = {Jamalaiah,  B.C. and Kumar,  J. Suresh and Babu,  A. Mohan and Moorthy,  L. Rama},
  year = {2009},
  month = jun,
  pages = {63--67}
}

@article{appl-inks,
  title = {Circularly polarized lanthanide luminescence for advanced security inks},
  volume = {5},
  ISSN = {2397-3358},
  DOI = {10.1038/s41570-020-00235-4},
  number = {2},
  journal = {Nature Reviews Chemistry},
  publisher = {Springer Science and Business Media LLC},
  author = {MacKenzie,  Lewis E. and Pal,  Robert},
  year = {2020},
  month = dec,
  pages = {109--124}
}

@article{Fe-ZrO2,
  title = {Optical absorption study of iron-substituted zirconia and yttria-stabilized zirconia using experimental measurements and many-body perturbation theory},
  volume = {8},
  ISSN = {2475-9953},
  DOI = {10.1103/physrevmaterials.8.085203},
  number = {8},
  journal = {Physical Review Materials},
  publisher = {American Physical Society (APS)},
  author = {Liu,  Shunshun and Champagne,  Victor K. and Clarke,  David R. and Balachandran,  Prasanna V.},
  year = {2024},
  month = aug 
}

@article{cryst-fiber,
  title = {{Er:CaF$_2$} single-crystal fiber {Q-switched} laser with diode pumping in the mid-infrared region},
  volume = {227},
  ISSN = {0022-2313},
  DOI = {10.1016/j.jlumin.2020.117519},
  journal = {Journal of Luminescence},
  publisher = {Elsevier BV},
  author = {Zong,  Mengyu and Yang,  Xiaojuan and Liu,  Jingjing and Zhang,  Zhen and Jiang,  Shouzhen and Liu,  Jie and Su,  Liangbi},
  year = {2020},
  month = nov,
  pages = {117519}
}

@article{Eu-big,
  title = "Interpretation of europium(III) spectra",
  journal="Coordination Chemistry Reviews",
  volume = "295",
  pages  = "1-45",
  year   = "2015",
  doi    = "10.1016/j.ccr.2015.02.015",
  author = "Binnemans, K"
}

@article{wan2,
  title = "Maximally localized Wannier functions: Theory and applications",
  journal= "Rev. Mod. Phys.",
  volume = "84",
  pages  = "1419",
  year   = "2012",
  doi    = "10.1103/RevModPhys.84.1419",
  author = "Marzari, N and Mostofi, A A and Yates, J R and Souza, I and Vanderbilt, D"
}

@article{wan1,
  title = "Maximally localized generalized Wannier functions for composite energy bands",
  journal= "Phys. Rev. B",
  volume = "56",
  pages  = "12847",
  year   = "1997",
  doi    = "10.1103/PhysRevB.56.12847",
  author = "Marzari, N and Vanderbilt, D"
}

@article{wan90,
  title = "Wannier90 as a community code: new features and applications",
  journal= "J. Phys. Cond. Mater.",
  volume = "32",
  pages  = "165902",
  year   = "2020",
  doi    = "10.1088/1361-648X/ab51ff",
  author = "Pizzi, G and Vitale, V and Arita, R and Bl\:ugel, S and Freimuth, F and G\'eranton, G and Gibertini, M and
Gresch, D and Johnson, C and Koretsune, T and Iba\~nez-Azpiroz, J and Lee, H and Lihm, J M and Marchand, D and Marrazzo, A and
Mokrousov, Y and Mustafa, J I and Nohara, Y and Nomura, Y and Paulatto, L and Ponc\'e, S and Ponweiser, T and Qiao, J and
Th\:ole, F and Tsirkin, S S and Wierzbowska, M and Marzari, N and Vanderbilt, D and Souza, I and Mostofi, A A and Yates, J R"
}

@article{MBPT,
  title = "New method for calculating the one-particle {Green's} function with application to the electron-gas problem",
  journal= "Phys. Rev.",
  volume = "139",
  pages  = "A796",
  year   = "1965",
  doi    = "10.1103/PhysRev.139.A796",
  author = "Hedin, L"
}

@article{BSE,
 title = "A Relativistic Equation for Bound-State Problems",
 journal = "Phys. Rev.",
 volume = "84",
 pages = "1232--1242",
 year = "1951",
 doi = "10.1103/PhysRev.84.1232",
 author = "Salpeter, E E and Bethe, H A"
}

@article{yambo-2,
  title = "Many-body perturbation theory calculations using the yambo code",
  journal= "J. Phys. Cond. Matter",
  volume = "31",
  pages  = "325902",
  year   = "2019",
  doi    = "10.1088/1361-648X/ab15d0",
  author = "Sangalli, D and Ferretti, A and Miranda, H and Attaccalite, C and Marri, I and Cannuccia, E and Melo, P M and Marsili, M and
Paleari, F and Marrazzo, A and Prandini, G and Bonf\'a, P and Atambo, M O and  Affinito, F and Palummo, M and Molina Sanchez, A 
and Hogan, C and Gr\:uning, M and Varsano, D and Marini, A"
}

@article{yambo-1,
  title = "Yambo: an ab initio tool for excited state calculations",
  journal= "Comp. Phys. Comm.",
  volume = "180",
  pages  = "1392",
  year   = "2009",
  doi    = "",
  author = "Marini, A and Hogan, C and Gr\:uning, M and Varsano, D"
}

@article{QE,
  title="QUANTUM ESPRESSO: a modular and open-source software project for quantum simulations of materials",
  journal="J. Phys.: Condens. Matter.",
  volume="21",
  pages="395502",
  year="2009",
  doi = "10.1088/0953-8984/21/39/395502",
  author="Giannozzi, P and Baroni, S and Bonini, M and Calandra, M and Car, R and Cavazzoni, C and Ceresoli, D and Chiarotti, G L and Cococcioni, M and Dabo, I and  Dal Corso, A and de Gironcoli, S and Fabris, S and Fratesi, G and Gebauer, R and Gerstmann, U and Gougoussis, C and Kokalj, A and Lazzeri, M and Martin-Samos, L and Marzari, N and Mauri, F and Mazzarello, R and Paolini, S and Pasquarello, A and Paulatto, L and Sbraccia, C and Scandolo, S and Sclauzero, G and Seitsonen, A P and Smogunov, A and Umari, P and Wentzcovitch, R M" 
}

@article{MP,
	title = "Special points for {Brillouin-zone} integrations",
	journal = "Phys. Rev. B",
	volume = "13",
	pages = "5188--5192",
        year = "1976",
        doi = "10.1103/PhysRevB.13.5188",
        author = "Monkhorst, Hendrik J. and Pack, James D."
}

@article{BFGS,
  title = "Practical Methods of Optimization",
  journal = "John Wiley and Sons, Ltd",
  volume = "sec. ed.",
  pages = "ISBN 978-3-319-35112-4",
  year="2000",
  doi = "10.1002/9781118723203",
  author="R. Fletcher"
}

@article{Sham, 
        author = {Kohn, W. and Sham, L. J.},
        journal = {Phys. Rev.},
        month = {Nov},
        pages = {A1133--A1138},
        title = {Self-Consistent Equations Including Exchange and Correlation Effects},
        volume = {140},
        year = {1965},
        doi = {10.1103/PhysRev.140.A1133}
}

@article{Kohn,
        author = {Hohenberg, P. and Kohn, W.},
        issue = {3B}, 
        journal = {Phys. Rev.},
        month = {Nov},
        pages = {B864--B871},
        title = {Inhomogeneous Electron Gas}, 
        volume = {136},
        year = {1964},
        doi = {10.1103/PhysRev.136.B864}
}

@article{PZ,
  title = {Self-interaction correction to density-functional approximations for many-electron systems},
  volume = {23},
  ISSN = {0163-1829},
  DOI = {10.1103/physrevb.23.5048},
  number = {10},
  journal = {Physical Review B},
  publisher = {American Physical Society (APS)},
  author = {Perdew,  J. P. and Zunger,  Alex},
  year = {1981},
  month = may,
  pages = {5048--5079}
}

@article{SOC-DFT,
        author = {Corso, Andrea Dal and Conte, Adriano Mosca},
        journal = {Phys. Rev. B},
        month = {Mar},
        pages = {115106},
        title = {Spin-orbit coupling with ultrasoft pseudopotentials: Application to {Au} and {Pt}},
        volume = {71},
        year = {2005},
        doi = {10.1103/PhysRevB.71.115106}
}

@article{SOC-BSE,
  title = {Spinorial formulation of the {GW-BSE} equations and spin properties of excitons in two-dimensional transition metal dichalcogenides},
  volume = {103},
  ISSN = {2469-9969},
  DOI = {10.1103/physrevb.103.155152},
  number = {15},
  journal = {Physical Review B},
  publisher = {American Physical Society (APS)},
  author = {Marsili,  Margherita and Molina-Sánchez,  Alejandro and Palummo,  Maurizia and Sangalli,  Davide and Marini,  Andrea},
  year = {2021},
  month = apr
}

@article{LDA+U-1,
  title = {Band theory and Mott insulators: {Hubbard U} instead of {Stoner I}},
  volume = {44},
  ISSN = {1095-3795},
  DOI = {10.1103/physrevb.44.943},
  number = {3},
  journal = {Physical Review B},
  publisher = {American Physical Society (APS)},
  author = {Anisimov,  Vladimir I. and Zaanen,  Jan and Andersen,  Ole K.},
  year = {1991},
  month = jul,
  pages = {943--954}
}

@article{LDA+U-2,
  title = {First-principles calculations of the electronic structure and spectra of strongly correlated systems: the {LDA+U} method},
  volume = {9},
  ISSN = {1361-648X},
  DOI = {10.1088/0953-8984/9/4/002},
  number = {4},
  journal = {Journal of Physics: Condensed Matter},
  publisher = {IOP Publishing},
  author = {Anisimov,  Vladimir I and Aryasetiawan,  F and Lichtenstein,  A I},
  year = {1997},
  month = jan,
  pages = {767--808}
}

@article{xcrysden,
  title = {{XCrySDen}\--a new program for displaying crystalline structures and electron densities},
  volume = {17},
  ISSN = {1093-3263},
  DOI = {10.1016/s1093-3263(99)00028-5},
  number = {3–4},
  journal = {Journal of Molecular Graphics and Modelling},
  publisher = {Elsevier BV},
  author = {Kokalj,  Anton},
  year = {1999},
  month = jun,
  pages = {176--179}
}

@article{hp,
  title = {{HP} $\--$ A code for the calculation of {Hubbard} parameters using density-functional perturbation theory},
  volume = {279},
  ISSN = {0010-4655},
  DOI = {10.1016/j.cpc.2022.108455},
  journal = {Computer Physics Communications},
  publisher = {Elsevier BV},
  author = {Timrov,  Iurii and Marzari,  Nicola and Cococcioni,  Matteo},
  year = {2022},
  month = oct,
  pages = {108455}
}

@article{hp1,
  author  = {Timrov, I. and Marzari, N. and Cococcioni, M.},
  title   = {Hubbard parameters from density-functional perturbation theory},
  journal = {Physical Review B},
  volume  = {98},
  number  = {8},
  pages   = {085127},
  numpages = {19},
  year    = {2018},
  month   = {Aug},
  publisher = {American Physical Society},
  doi     = {10.1103/PhysRevB.98.085127},
}

@article{hp2,
  author    = {Timrov, I. and Marzari, N. and Cococcioni, M.},
  title     = {Self-consistent Hubbard parameters from density-functional perturbation theory in the ultrasoft and projector-augmented wave formulations},
  journal   = {Physical Review B},
  volume    = {103},
  number    = {4},
  pages     = {045141},
  numpages  = {23},
  year      = {2021},
  month     = {Jan},
  publisher = {American Physical Society},
  doi       = {10.1103/PhysRevB.103.045141},
}

@article{matteo-2005,
  author    = {Cococcioni, Matteo and de Gironcoli, Stefano},
  title     = {Linear response approach to the calculation of the effective interaction parameters in the $\mathrm{LDA+U}$ method},
  journal   = {Physical Review B},
  volume    = {71},
  number    = {3},
  pages     = {035105},
  numpages  = {19},
  year      = {2005},
  month     = {Jan},
  publisher = {American Physical Society},
  doi       = {10.1103/PhysRevB.71.035105},
}

@article{UBS1,
  author    = {Mossaenek, R. J. O. and Abbott, M. and Schlueter, C. and Lee, T.-L. and Sing, M. and Claessen, R. and Thiess, S.},
  title     = {Enhanced electron correlations at the surface of $\mathrm{Sr}_x\mathrm{Ca}_{1-x}\mathrm{VO}_3$},
  journal   = {Physical Review B},
  volume    = {91},
  number    = {16},
  pages     = {165123},
  year      = {2015},
  month     = {Apr},
  publisher = {American Physical Society},
  doi       = {10.1103/PhysRevB.91.165123},
}

@article{UBS2,
  author    = {Maiti, Kalobaran and Singh, Ravi S. and Medicherla, V. R. R.},
  title     = {Surface and bulk electronic structure of $\mathrm{Sr}_2\mathrm{RuO}_4$: Evidence for electronic inhomogeneity},
  journal   = {Physical Review B},
  volume    = {70},
  number    = {19},
  pages     = {195112},
  year      = {2004},
  month     = {Nov},
  publisher = {American Physical Society},
  doi       = {10.1103/PhysRevB.70.195112},
}

@article{UBS3,
  author    = {Kabir, M.},
  title     = {Self-consistent determination of {Hubbard $U$} for explaining the anomalous magnetic moment of $\mathrm{Gd}_{13}$ cluster},
  journal   = {Physical Review B},
  volume    = {89},
  number    = {8},
  pages     = {085103},
  year      = {2014},
  month     = {Feb},
  publisher = {American Physical Society},
  doi       = {10.1103/PhysRevB.89.085103},
}

@article{UBS4,
  author  = {Hu, Wende and Cao, Xiao-Ming and Hu, P.},
  title   = {{DFT+U} Study on Catalysis by $\mathrm{Co_3O_4}$: Influence of {$U$} Value and a Surface--Bulk Bi-{$U$} Strategy},
  journal = {The Journal of Physical Chemistry C},
  volume  = {122},
  number  = {34},
  pages   = {19593--19602},
  year    = {2018},
  doi     = {10.1021/acs.jpcc.8b05513},
}

@article{U-different,
  title = {Extensive Benchmarking of {DFT+U} Calculations for Predicting Band Gaps},
  volume = {11},
  ISSN = {2076-3417},
  DOI = {10.3390/app11052395},
  number = {5},
  journal = {Applied Sciences},
  publisher = {MDPI AG},
  author = {Kirchner-Hall,  Nicole E. and Zhao,  Wayne and Xiong,  Yihuang and Timrov,  Iurii and Dabo,  Ismaila},
  year = {2021},
  month = Mar,
  pages = {2395}
}

@article{U-China,
  title = {Electronic structures and optical properties of monoclinic ZrO$_2$ studied by first-principles local density approximation {+ U} approach},
  volume = {6},
  ISSN = {2227-8508},
  DOI = {10.1007/s40145-016-0216-y},
  number = {1},
  journal = {Journal of Advanced Ceramics},
  publisher = {Tsinghua University Press},
  author = {Li,  Jinping and Meng,  Songhe and Niu,  Jiahong and Lu,  Hantao},
  year = {2017},
  month = Mar,
  pages = {43–49}
}

@misc{yambopy,
  author = {YAMBO}, 
  title = {Yambopy tutorial: {Yambo} databases},
  howpublished = {\url{http://wiki.yambo-code.eu/wiki/index.php/Yambopy_tutorial:_Yambo_databases}},
  year         = {2024},
  note         = {Accessed: 2026-03-10}
}

@article{charge,
  title = {Electronic structure and total-energy migration barriers of silicon self-interstitials},
  volume = {30},
  ISSN = {0163-1829},
  DOI = {10.1103/physrevb.30.1844},
  number = {4},
  journal = {Physical Review B},
  publisher = {American Physical Society (APS)},
  author = {Bar-Yam,  Y. and Joannopoulos,  J. D.},
  year = {1984},
  month = aug,
  pages = {1844--1852}
}

@article{JO-1,
  title = {Optical Absorption Intensities of Rare-Earth Ions},
  volume = {127},
  ISSN = {0031-899X},
  DOI = {10.1103/physrev.127.750},
  number = {3},
  journal = {Physical Review},
  publisher = {American Physical Society (APS)},
  author = {Judd,  B. R.},
  year = {1962},
  month = aug,
  pages = {750–761}
}

@article{JO-2,
  title = {Intensities of Crystal Spectra of Rare-Earth Ions},
  volume = {37},
  ISSN = {1089-7690},
  DOI = {10.1063/1.1701366},
  number = {3},
  journal = {The Journal of Chemical Physics},
  publisher = {AIP Publishing},
  author = {Ofelt,  G. S.},
  year = {1962},
  month = aug,
  pages = {511–520}
}

@article{Brik-model,
  title = {Spectral analysis of {Er$^{3+}$-},  {Er$^{3+}$/Yb$^{3+}$-} and {Er$^{3+}$/Tm$^{3+}$/Yb$^{3+}$-}doped {TeO$_2$\--ZnO\--WO$_3$\--TiO$_2$\--Na$_2$O} glasses},
  volume = {20},
  ISSN = {1361-648X},
  DOI = {10.1088/0953-8984/20/37/375101},
  number = {37},
  journal = {Journal of Physics: Condensed Matter},
  publisher = {IOP Publishing},
  author = {Lakshminarayana,  G and Qiu,  Jianrong and Brik,  M G and Kumar,  G A and Kityk,  I V},
  year = {2008},
  month = aug,
  pages = {375101}
}

@inproceedings{temp-2,
  title = {Optical thermometry using temperature-dependent absorption in rare earth-doped crystals},
  DOI = {10.1117/12.3042382},
  booktitle = {Optical Components and Materials XXII},
  publisher = {SPIE},
  author = {Dwyer,  Ashley E. and Ganem,  Joseph},
  editor = {Digonnet,  Michel J. and Jiang,  Shibin},
  year = {2025},
  month = mar,
  pages = {6}
}

@article{temp-3,
  title = {Optical thermometry based on temperature-dependent luminescence of {Eu$^{2+}$/Tb$^{3+}$/Eu$^{3+}$} {Co}-doped {BaO-ZnO-P$_2$O$_5$-B$_2$O$_3$} glasses},
  volume = {51},
  ISSN = {0272-8842},
  DOI = {10.1016/j.ceramint.2025.04.299},
  number = {20},
  journal = {Ceramics International},
  publisher = {Elsevier BV},
  author = {Wang,  Siping and Chen,  Zhaoyang and Bi,  Shala and Gao,  Zuyang and Hu,  Houjin and Liu,  Yi and Zhang,  Bolong},
  year = {2025},
  month = aug,
  pages = {31059–31068}
}

@article{temp-4,
  title = {Temperature dependence of the luminescence spectra of garnet crystals doped with chromium ions},
  volume = {31},
  ISSN = {0925-3467},
  DOI = {10.1016/j.optmat.2007.12.016},
  number = {7},
  journal = {Optical Materials},
  publisher = {Elsevier BV},
  author = {Orucu,  Humeyra and Ozen,  Gonul and Collins,  John and Di Bartolo,  Baldassare},
  year = {2009},
  month = may,
  pages = {1065–1070}
}

@article{f-mol,
  title = {The design of responsive luminescent lanthanide probes and sensors},
  volume = {50},
  ISSN = {1460-4744},
  DOI = {10.1039/d1cs00310k},
  number = {14},
  journal = {Chemical Society Reviews},
  publisher = {Royal Society of Chemistry (RSC)},
  author = {Parker,  David and Fradgley,  Jack D. and Wong,  Ka-Leung},
  year = {2021},
  pages = {8193–8213}
}

@article{heavy-scintillator,
  title = {Single crystal growth and luminescent properties of {YSH:Eu} scintillator by optical floating zone method},
  volume = {738},
  ISSN = {0009-2614},
  DOI = {10.1016/j.cplett.2019.136916},
  journal = {Chemical Physics Letters},
  publisher = {Elsevier BV},
  author = {Yu,  Haihang and Liu,  Changjiang and Zhang,  Zhikun and Huang,  Shimin and Yang,  Yunling and Mao,  Rihua and Feng,  He and Zhao,  Jingtai},
  year = {2020},
  month = jan,
  pages = {136916}
}

@article{Eu2BaBrI,
  title = {Luminescence of {BaBrI} and {SrBrI} single crystals doped with {Eu$^{2+}$}},
  volume = {79},
  ISSN = {0925-3467},
  DOI = {10.1016/j.optmat.2018.03.017},
  journal = {Optical Materials},
  publisher = {Elsevier BV},
  author = {Shalaev,  A.A. and Shendrik,  R. and Myasnikova,  A.S. and Bogdanov,  A. and Rusakov,  A. and Vasilkovskyi,  A.},
  year = {2018},
  month = may,
  pages = {84–89}
}

@article{medicine,
  title = {Recent advances in synthesis and surface modification of lanthanide-doped upconversion nanoparticles for biomedical applications},
  volume = {30},
  ISSN = {0734-9750},
  DOI = {10.1016/j.biotechadv.2012.04.009},
  number = {6},
  journal = {Biotechnology Advances},
  publisher = {Elsevier BV},
  author = {Lin,  Min and Zhao,  Ying and Wang,  ShuQi and Liu,  Ming and Duan,  ZhenFeng and Chen,  YongMei and Li,  Fei and Xu,  Feng and Lu,  TianJian},
  year = {2012},
  month = nov,
  pages = {1551–1561}
}

@article{O-charge,
  title = {Unraveling the Charge State of Oxygen Vacancies in {ZrO$_{2-x}$} on the Basis of Synergistic Computational and Experimental Evidence},
  volume = {123},
  ISSN = {1932-7455},
  DOI = {10.1021/acs.jpcc.9b00411},
  number = {18},
  journal = {The Journal of Physical Chemistry C},
  publisher = {American Chemical Society (ACS)},
  author = {Imparato,  Claudio and Fantauzzi,  Marzia and Passiu,  Cristiana and Rea,  Ilaria and Ricca,  Chiara and Aschauer,  Ulrich and Sannino,  Filomena and D’Errico,  Gerardino and De Stefano,  Luca and Rossi,  Antonella and Aronne,  Antonio},
  year = {2019},
  month = mar,
  pages = {11581–11590}
}

@article{Y2O3-tZrO2,
  title = {Influence of oxygen vacancy compensation on the structure,  electronic and mechanical properties of yttrium stabilized tetragonal zirconia},
  volume = {135},
  ISSN = {1369-8001},
  DOI = {10.1016/j.mssp.2021.106082},
  journal = {Materials Science in Semiconductor Processing},
  publisher = {Elsevier BV},
  author = {Fan,  Zhou and Wang,  Yang and Zhang,  Yidong and Liu,  Jianyi},
  year = {2021},
  month = nov,
  pages = {106082}
}

@article{O-trap,
  title = {Effect of Oxygen Vacancy on Phase Transition and Photoluminescence Properties of Nanocrystalline Zirconia Synthesized by the One-Pot Reaction},
  volume = {113},
  ISSN = {1932-7455},
  DOI = {10.1021/jp8103497},
  number = {31},
  journal = {The Journal of Physical Chemistry C},
  publisher = {American Chemical Society (ACS)},
  author = {Cong,  Yan and Li,  Bin and Yue,  Shumei and Fan,  Di and Wang,  Xiao-jun},
  year = {2009},
  pages = {13974–13978}
}

@article{O-photocatalysis,
  title = {Oxygen vacancy-rich mesoporous {ZrO$_2$} with remarkably enhanced visible-light photocatalytic performance},
  volume = {182},
  ISSN = {0927-0248},
  DOI = {10.1016/j.solmat.2018.03.023},
  journal = {Solar Energy Materials and Solar Cells},
  publisher = {Elsevier BV},
  author = {Zhang,  Jingji and Gao,  Yafeng and Jia,  Xuanrui and Wang,  Jiangying and Chen,  Zhi and Xu,  Yu},
  year = {2018},
  month = aug,
  pages = {113–120}
}

@article{Fabris,
  title = {A stabilization mechanism of zirconia based on oxygen vacancies only},
  volume = {50},
  ISSN = {1359-6454},
  DOI = {10.1016/s1359-6454(02)00385-3},
  number = {20},
  journal = {Acta Materialia},
  publisher = {Elsevier BV},
  author = {Fabris,  S},
  year = {2002},
  month = dec,
  pages = {5171–5178}
}

@article{defects,
  title = {Photoluminescence of Tetragonal {ZrO$_2$} Nanoparticles: Contribution from $Oh$-Groups,  Oxygen Vacancies,  and Impurities},
  DOI = {10.2139/ssrn.5059551},
  publisher = {Elsevier BV},
  journal = {SSRN},
  author = {Kostyukov,  Anton and Nashivochnikov,  Aleksandr and Rakhmanova,  Mariana and Panchenko,  Valentina and Pochtar’,  Alyona  A. and Cherepanova,  Svetlana  V. and Snytnikov,  Valeriy  N.},
  year = {2024}
}

@article{solar-PV,
  title = {Oxygen-Deficient Zirconia {(ZrO$_{2-x}$):} A New Material for Solar Light Absorption},
  volume = {6},
  ISSN = {2045-2322},
  DOI = {10.1038/srep27218},
  number = {1},
  journal = {Scientific Reports},
  publisher = {Springer Science and Business Media LLC},
  author = {Sinhamahapatra,  Apurba and Jeon,  Jong-Pil and Kang,  Joonhee and Han,  Byungchan and Yu,  Jong-Sung},
  year = {2016},
  month = jun 
}

@article{hydrolysis,
  title = {Synthesis of nanostructured tetragonal {ZrO$_2$} of enhanced thermal stability},
  DOI = {},
  volume = {51},
  pages = {S27-S33},
  journal = {Nukleonika},
  author = {Adamski, Andrzej and Jakubus, Pawe\l{} and Sojka, Zbigniew},
  year = {2006}
}

@article{glass-2024,
  title = {Structural and luminescence properties of novel {Eu$^{3+}$}-doped Na3Ba2LaNb10O30 phosphors with high quantum efficiency and excellent color purity for w-LED applications},
  volume = {14},
  ISSN = {2046-2069},
  DOI = {10.1039/d4ra05026f},
  number = {40},
  journal = {RSC Advances},
  publisher = {Royal Society of Chemistry (RSC)},
  author = {Poria,  Kanishk and Bhatia,  Sanjana and Kashyap,  Rajiv and Kashyap,  Vikas and Sihmar,  Isha and Deopa,  Nisha and Shahi,  Jangvir Singh},
  year = {2024},
  pages = {29490–29504}
}

@article{sol-gel,
  title = {Sol\--gel synthesis of tetragonal {ZrO$_2$} nanoparticles stabilized by crystallite size and oxygen vacancies},
  volume = {67},
  ISSN = {1573-4846},
  DOI = {10.1007/s10971-013-3112-8},
  number = {3},
  journal = {Journal of Sol-Gel Science and Technology},
  publisher = {Springer Science and Business Media LLC},
  author = {Mahmood,  Qaiser and Afzal,  Adeel and Siddiqi,  Humaira M. and Habib,  Amir},
  year = {2013},
  month = jul,
  pages = {670–674}
}

@article{geom-1,
  title = {Tetragonal\--Monoclinic Phase Transition Enthalpy and Temperature of {ZrO$_2$-CeO$_2$} Solid Solutions},
  volume = {78},
  ISSN = {1551-2916},
  DOI = {10.1111/j.1151-2916.1995.tb08642.x},
  number = {8},
  journal = {Journal of the American Ceramic Society},
  publisher = {Wiley},
  author = {Yashima,  Masatomo and Mitsuhashi,  Takefumi and Takashina,  Hiroyuki and Kakihana,  Masato and Ikegami,  Takayasu and Yoshimura,  Masahiro},
  year = {1995},
  month = aug,
  pages = {2225–2228}
}

@article{geom-2,
  title = {Thermodynamic Assessment of the {ZrO$_2$-YO$_{1.5}$} System},
  volume = {74},
  ISSN = {1551-2916},
  DOI = {10.1111/j.1151-2916.1991.tb07142.x},
  number = {7},
  journal = {Journal of the American Ceramic Society},
  publisher = {Wiley},
  author = {Du,  Yong and Jin,  Zhanpeng and Huang,  Peiyun},
  year = {1991},
  month = jul,
  pages = {1569–1577}
}

@article{geom-3,
  title = {Structural changes of {ZrO$_2$-CeO$_2$} solid solutions around the monoclinic-tetragonal phase boundary},
  volume = {51},
  ISSN = {1095-3795},
  DOI = {10.1103/physrevb.51.8018},
  number = {13},
  journal = {Physical Review B},
  publisher = {American Physical Society (APS)},
  author = {Yashima,  Masatomo and Hirose,  Teruo and Katano,  Susumu and Suzuki,  Yasuo and Kakihana,  Masato and Yoshimura,  Masahiro},
  year = {1995},
  month = apr,
  pages = {8018–8025}
}

@article{theor-exp-geom,
  title = {Study of {ZrO$_2$} phase structure and electronic properties},
  volume = {34},
  ISSN = {1029-0435},
  DOI = {10.1080/08927020802101759},
  number = {10–15},
  journal = {Molecular Simulation},
  publisher = {Informa UK Limited},
  author = {Qunbo,  Fan and Fuchi,  Wang and Huiling,  Zhang and Feng,  Zhang},
  year = {2008},
  month = sep,
  pages = {1099–1103}
}

@article{1994,
  title = {Experimental and theoretical determination of the electronic structure and optical properties od three phases of {ZrO$_2$}},
  author = {French, R. H. and Glass, S. J. and Ohuchi, F. S. and Xu, Y.-N. and Ching, W. Y.},
  volume = {49},
  number = {8},
  year = {1994},
  journal = {Physical Review B},
  month = feb,
  pages = {5133--5142}
}

@article{U-calc-Eu,
  doi = {10.48550/ARXIV.2211.16260},
  author = {Liu,  Bei-Lei and Wang,  Yue-Chao and Liu,  Yu and Xu,  Yuan-Ji and Chen,  Xin and Song,  Hong-Zhou and Bi,  Yan and Liu,  Hai-Feng and Song,  Hai-Feng},
  title = {Comparative study of first-principles approaches for effective {Coulomb} interaction strength {$U_{eff}$} between localized $f$-electrons: lanthanide metals as an example},
  journal = {arXiv},
  year = {2022},
  copyright = {arXiv.org perpetual,  non-exclusive license}
}

@article{U-calc-ZrO2,
  title = {Band Structure Engineering and Optical Properties of Pristine and Doped Monoclinic Zirconia {(m-ZrO$_2$)}: Density Functional Theory Theoretical Prospective},
  volume = {6},
  ISSN = {2470-1343},
  DOI = {10.1021/acsomega.1c04756},
  number = {44},
  journal = {ACS Omega},
  publisher = {American Chemical Society (ACS)},
  author = {Khattab,  El-Sayed R. and Abd El Rehim,  Sayed S. and Hassan,  Walid M. I. and El-Shazly,  Tamer S.},
  year = {2021},
  month = oct,
  pages = {30061–30068}
}

@article{Ricca,
  title = {Revealing the properties of the cubic {ZrO$_2$} (111) surface by periodic DFT calculations: reducibility and stabilization through doping with aliovalent {Y$_2$O$_3$}},
  volume = {5},
  ISSN = {2046-2069},
  DOI = {10.1039/c4ra15206a},
  number = {18},
  journal = {RSC Advances},
  publisher = {Royal Society of Chemistry (RSC)},
  author = {Ricca,  Chiara and Ringued/'e,  Armelle and Cassir,  Michel and Adamo,  Carlo and Labat,  Fr/'ed/'eric},
  year = {2015},
  pages = {13941–13951}
}

@article{O-blue,
  title = {Oxygen vacancies in zirconium oxide as the blue luminescence centres and traps responsible for charge transport: Part I\--Crystals},
  volume = {15},
  ISSN = {2589-1529},
  DOI = {10.1016/j.mtla.2020.100979},
  journal = {Materialia},
  publisher = {Elsevier BV},
  author = {Islamov,  Damir R. and Gritsenko,  Vladimir A. and Perevalov,  Timofey V. and Yelisseyev,  Alexander P. and Pustovarov,  Vladimir A. and Korolkov,  Ilya V. and Lomonova,  Elena E.},
  year = {2021},
  month = mar,
  pages = {100979}
}

@article{lumin-ff-ZrO2,
  title = {Photoluminescence properties of {Eu$^{3+}$} doped {ZrO$_2$} with different morphologies and crystal structures},
  volume = {864},
  ISSN = {0925-8388},
  DOI = {10.1016/j.jallcom.2021.158781},
  journal = {Journal of Alloys and Compounds},
  publisher = {Elsevier BV},
  author = {Liu,  Dongxin and Wang,  Mingli and Gong,  Liyuan and Zhao,  Jianling and Zhu,  Miao and Wang,  Xixin},
  year = {2021},
  month = may,
  pages = {158781}
}

@article{morphology,
  title = {Photoluminescence properties of {Eu$^{3+}$} doped {ZrO$_2$} with different morphologies and crystal structures},
  volume = {864},
  ISSN = {0925-8388},
  DOI = {10.1016/j.jallcom.2021.158781},
  journal = {Journal of Alloys and Compounds},
  publisher = {Elsevier BV},
  author = {Liu,  Dongxin and Wang,  Mingli and Gong,  Liyuan and Zhao,  Jianling and Zhu,  Miao and Wang,  Xixin},
  year = {2021},
  month = may,
  pages = {158781}
}

@article{from-Witek,
  title = {Effect of the crystalline structure and the oxidation state on the luminescence quenching in {ZrO$_2$:Ln$^{3+}$} {(Ln = Pr,  Nd,  Eu,  Gd,  Dy,  Yb)}},
  volume = {1022},
  ISSN = {0925-8388},
  DOI = {10.1016/j.jallcom.2025.179989},
  journal = {Journal of Alloys and Compounds},
  publisher = {Elsevier BV},
  author = {Carrera-Peralta,  Rosalba and L/'opez-Pacheco,  Germ/'an and Gonz/'alez,  Federico},
  year = {2025},
  month = apr,
  pages = {179989}
}

@article{anti-counterfeiting,
  title = {Phase and photoluminescence analysis of dual-color emissive {Eu$^{3+}$}-doped {ZrO$_2$} nanoparticles for advanced security features in anti-counterfeiting},
  volume = {631},
  ISSN = {0927-7757},
  DOI = {10.1016/j.colsurfa.2021.127715},
  journal = {Colloids and Surfaces A: Physicochemical and Engineering Aspects},
  publisher = {Elsevier BV},
  author = {King,  Abhishek and Singh,  Rahul and Nayak,  Bibhuti B.},
  year = {2021},
  month = dec,
  pages = {127715}
}

@article{YZO,
  title = {The crystal structure of tetragonal {ZrO$_2$}},
  volume = {15},
  ISSN = {0365-110X},
  DOI = {10.1107/s0365110x62003114},
  number = {11},
  journal = {Acta Crystallographica},
  publisher = {International Union of Crystallography (IUCr)},
  author = {Teufer,  G.},
  year = {1962},
  month = nov,
  pages = {1187–1187}
}

@article{MD-glass,
  title = {Molecular-dynamics modeling of {Eu$^{3+}$}-ion clustering in {SiO$_2$} glass},
  volume = {79},
  ISSN = {1550-235X},
  DOI = {10.1103/physrevb.79.024202},
  number = {2},
  journal = {Physical Review B},
  publisher = {American Physical Society (APS)},
  author = {Afify,  N. D. and Mountjoy,  G.},
  year = {2009},
  month = jan 
}

@article{JO-ff,
  title = {Blue emission and energy transfer characteristics of {Eu-}doped sol\--gel glasses without post-processing},
  volume = {293–295},
  ISSN = {0022-3093},
  DOI = {10.1016/s0022-3093(01)00756-6},
  journal = {Journal of Non-Crystalline Solids},
  publisher = {Elsevier BV},
  author = {Kamata,  N and Satoh,  C and Tosaka,  K and Yamada,  K},
  year = {2001},
  month = nov,
  pages = {595–599}
}

@article{Tauc,
  title = {Optical and structural properties of yttrium doped zinc lead borate glasses},
  volume = {274},
  ISSN = {0030-4026},
  DOI = {10.1016/j.ijleo.2023.170542},
  journal = {Optik},
  publisher = {Elsevier BV},
  author = {Ahmed,  E.M. and Elzelaky,  A.A. and El-Ghamaz,  N.A. and Youssif,  M.I.},
  year = {2023},
  month = mar,
  pages = {170542}
}

@article{ASF-1,
  title = {A new method for the determination of optical band gap and the nature of optical transitions in semiconductors},
  volume = {119},
  ISSN = {1432-0649},
  DOI = {10.1007/s00340-015-6053-9},
  number = {2},
  journal = {Applied Physics B},
  publisher = {Springer Science and Business Media LLC},
  author = {Souri,  Dariush and Tahan,  Zahra Esmaeili},
  year = {2015},
  month = feb,
  pages = {273–279}
}

@article{ASF-2,
  title = {An alternative procedure for the determination of the optical band gap and thickness of amorphous carbon nitride thin films},
  volume = {254},
  ISSN = {0169-4332},
  DOI = {10.1016/j.apsusc.2007.07.052},
  number = {1},
  journal = {Applied Surface Science},
  publisher = {Elsevier BV},
  author = {Escobar-Alarcón,  L. and Arrieta,  A. and Camps,  E. and Muhl,  S. and Rodil,  S. and Vigueras-Santiago,  E.},
  year = {2007},
  month = oct,
  pages = {412–415}
}

@article{HEM-1,
  title = {Determination of the optical constants of zinc oxide thin films by spectroscopic ellipsometry},
  volume = {72},
  ISSN = {1077-3118},
  DOI = {10.1063/1.121617},
  number = {25},
  journal = {Applied Physics Letters},
  publisher = {AIP Publishing},
  author = {Washington,  P. L. and Ong,  H. C. and Dai,  J. Y. and Chang,  R. P. H.},
  year = {1998},
  month = jun,
  pages = {3261–3263}
}

@article{HEM-2,
  title = {Physical,  structural and optical characterization of silicate modified bismuth-borate-tellurite glasses},
  volume = {1127},
  ISSN = {0022-2860},
  DOI = {10.1016/j.molstruc.2016.08.033},
  journal = {Journal of Molecular Structure},
  publisher = {Elsevier BV},
  author = {Berwal,  Neelam and Dhankhar,  Sunil and Sharma,  Preeti and Kundu,  R.S. and Punia,  R. and Kishore,  N.},
  year = {2017},
  month = jan,
  pages = {636–644}
}

@article{Elliott,
  title = {Intensity of Optical Absorption by Excitons},
  volume = {108},
  ISSN = {0031-899X},
  DOI = {10.1103/physrev.108.1384},
  number = {6},
  journal = {Physical Review},
  publisher = {American Physical Society (APS)},
  author = {Elliott,  R. J.},
  year = {1957},
  month = dec,
  pages = {1384–1389}
}

@article{laser-model,
  title = {Distributed Model for Actively {Q-Switched} Erbium-Doped Fiber Lasers},
  volume = {47},
  ISSN = {1558-1713},
  DOI = {10.1109/jqe.2011.2143695},
  number = {7},
  journal = {IEEE Journal of Quantum Electronics},
  publisher = {Institute of Electrical and Electronics Engineers (IEEE)},
  author = {Kolpakov,  Stanislav A. and Barmenkov,  Yuri O. and Guzman-Chavez,  Ana Dinora and Kir’yanov,  Alexander V. and Cruz,  José Luis and Diez,  Antonio and Andres,  Miguel V.},
  year = {2011},
  month = jul,
  pages = {928–934}
}

@article{high-power,
  title = {Pulsed Regimes of Erbium-Doped Fiber Laser {Q-}Switched Using Acousto-Optical Modulator},
  volume = {20},
  ISSN = {1558-4542},
  DOI = {10.1109/jstqe.2014.2304423},
  number = {5},
  journal = {IEEE Journal of Selected Topics in Quantum Electronics},
  publisher = {Institute of Electrical and Electronics Engineers (IEEE)},
  author = {Barmenkov,  Yuri O. and Kir’yanov,  Alexander V. and Cruz,  Jose L. and Andres,  Miguel V.},
  year = {2014},
  month = sep,
  pages = {337–344}
}

@article{fibre-nasi,
  title = {Fibre lasers - conditioning constructional and technological},
  volume = {58},
  ISSN = {0239-7528},
  DOI = {10.2478/v10175-010-0048-9},
  number = {4},
  journal = {Bulletin of the Polish Academy of Sciences: Technical Sciences},
  publisher = {Polish Academy of Sciences Chancellery},
  author = {Zając,  A. and Dorosz,  D. and Kochanowicz,  M. and Skórczakowski,  M. and Świderski,  J.},
  year = {2010},
  month = jan
}

@article{cluster,
  title = {Structure of phosphate and iron-phosphate glasses by {DFT} calculations and {FTIR/Raman} spectroscopy},
  volume = {450},
  ISSN = {0022-3093},
  DOI = {10.1016/j.jnoncrysol.2016.07.027},
  journal = {Journal of Non-Crystalline Solids},
  publisher = {Elsevier BV},
  author = {Stoch,  Pawel and Stoch,  Agata and Ciecinska,  Malgorzata and Krakowiak,  Ireneusz and Sitarz,  Maciej},
  year = {2016},
  month = oct,
  pages = {48–60}
}

@article{exp-Borik,
  title = {Optical spectroscopy of {ZrO2-Eu2O3} tetragonal solid solutions with monoclinic phase inclusions},
  volume = {1037},
  ISSN = {0925-8388},
  DOI = {10.1016/j.jallcom.2025.182340},
  journal = {Journal of Alloys and Compounds},
  publisher = {Elsevier BV},
  author = {Borik,  M.A. and Bukarev,  S.A. and Gerasimov,  M.V. and Kulebyakin,  A.V. and Lomonova,  E.E. and Myzina,  V.A. and Reu,  A.A. and Ryabochkina,  P.A. and Tabachkova,  N.Yu. and Volkova,  T.V.},
  year = {2025},
  month = Aug,
  pages = {182340}
}

@article{chemisensors,
  title = {Novel Photoluminescence and Optical Thermometry of Solvothermally Derived Tetragonal ZrO2:Ti4+, Eu3+ Nanocrystals},
  volume = {12},
  ISSN = {2227-9040},
  DOI = {10.3390/chemosensors12040062},
  number = {4},
  journal = {Chemosensors},
  publisher = {MDPI AG},
  author = {Li,  Lu and Qu,  Xuesong and Pan,  Guo-Hui and Jeong,  Jung Hyun},
  year = {2024},
  month = Apr,
  pages = {62}
}

@article{exp-jalcom,
  title = {Photoluminescence properties of Eu3+ doped ZrO2 with different morphologies and crystal structures},
  volume = {864},
  ISSN = {0925-8388},
  DOI = {10.1016/j.jallcom.2021.158781},
  journal = {Journal of Alloys and Compounds},
  publisher = {Elsevier BV},
  author = {Liu,  Dongxin and Wang,  Mingli and Gong,  Liyuan and Zhao,  Jianling and Zhu,  Miao and Wang,  Xixin},
  year = {2021},
  month = {May},
  pages = {158781}
}

\end{document}